\documentclass[a4paper,11pt]{article}
\pdfoutput=1
\usepackage{aas_macros}
\usepackage{jcappub} 
\usepackage{graphicx}
\graphicspath{{figures/}}
\usepackage{natbib}
\usepackage[colorlinks=true]{hyperref}

\def\tegr{{\theta_{E,\textrm{GR}}}}
\def\gppn{{\gamma_\textrm{PPN}}}

\title{\boldmath Test of the Equivalence Principle in the Dark Sector on Galactic Scales}
\author[a]{N. Mohapi,}
\author[a]{A. Hees,}
\author[a]{J. Larena}

\affiliation[a]{Department of Mathematics, Rhodes University,\\
Grahamstown, 6140, South Africa}

\emailAdd{n.mohapi@gmail.com}
\emailAdd{a.hees@ru.ac.za}
\emailAdd{j.larena@ru.ac.za}

\abstract{The Einstein Equivalence Principle is a fundamental principle of the theory of General Relativity. While this principle has been thoroughly tested with standard matter, the question of its validity in the Dark sector remains open. In this paper, we consider a general tensor-scalar theory that allows to test the equivalence principle in the Dark sector by introducing two different conformal couplings to standard matter and to Dark matter. We constrain these couplings by considering galactic observations of strong lensing and of velocity dispersion. Our analysis shows that, in the case of a violation of the Einstein Equivalence Principle, data favour violations through coupling strengths that are of opposite signs for ordinary and Dark matter. At the same time, our analysis does not show any significant deviations from General Relativity.}

\begin{document}
\maketitle
\flushbottom
\section{Introduction} \label{Intro}

The classical theory of General Relativity (GR) is the current paradigm to describe the gravitational interaction. So far, it has passed all the tests with flying colours~\cite{cliffwill,will:2014la}. Nevertheless, despite its successes, there are strong indications that GR is not the ultimate theory of gravitation. First of all, a quantum theory of gravitation is required to understand processes
happening in very strong gravitational fields like the ones in the early
Universe. Furthermore, it is often believed that GR and the standard model of particles are only approximations of a more fundamental unified theory. Finally, the observations that require the introduction of Dark Matter (DM) and Dark Energy are sometimes interpreted as hints towards a modification of the theory of gravitation on large scales. It is therefore crucial to search for any deviations from GR and to constrain them.

GR is built upon two principles: (i) the Einstein Equivalence Principle (EEP) and (ii) the Einstein field equations. The first principle gives a geometric nature to gravitation by identifying this interaction with space-time curvature parametrized by a space-time metric $g_{\mu\nu}$ \cite{cliffwill,will:2014la}. More precisely, it stipulates that there exists a metric to which all matter is minimally coupled to~\cite{thorne:1973fk}. This allows one to write the matter part of the action as
\begin{align}
	S_\textrm{mat}=\int d^4x \sqrt{-g}\mathcal L_\textrm{mat}(g_{\mu\nu},\Psi)\, ,
\end{align} 
where $g$ is the determinant of the space-time metric $g_{\mu\nu}$, $\mathcal L_\textrm{mat}$ is the matter Lagrangian and $\Psi$ denotes the matter fields. The second principle, the Einstein field equations, specifies the form of this metric, which in GR is determined by solving the field equations. These two principles have been thoroughly tested.

From a phenomenological point of view, three aspects of the EEP can be tested \cite{cliffwill,will:2014la}: (i) the Universality of Free Fall (UFF), (ii) the Local Lorentz Invariance (LLI) and (iii) the Local Position Invariance (LPI). The UFF states that the motion of a test body is independent of its composition. It has been constrained with various experiments. Precise tests of the UFF compare the free fall accelerations, $a_1$ and $a_2$, of two  different test bodies $1$ and $2$ falling in the gravitational field sourced by a body $S$. A succinct expression for the test of the UFF takes the form~\cite{cliffwill,will:2014la}
\begin{equation} \label{eq:UFF}
\left(\frac{\Delta a}{a}\right)_{S;1-2} =2\frac{a_1-a_2}{a_1+a_2} \approx \left(\frac{m_P}{m_A}\right)_{1} - \left(\frac{m_P}{m_A}\right)_{2}\, ,
\end{equation}
where $m_P$ and $m_A$ are the passive and active masses of each body. The UFF has been tested in various cases. First, it has been tested by comparing the motion of two different macroscopic bodies of standard luminous matter (SM). The best current constraints on the relative differential acceleration of two bodies are at the level of $10^{-13}$ and have been obtained by using Lunar Laser Ranging observations~\cite{williams:2009ys,williams:2012zr} and torsion balances~\cite{schlamminger:2008zr,adelberger:2009fk}. More recently, the UFF has been tested by comparing the acceleration measured by a macroscopic mass with the acceleration measured by a microscopic, quantum system and also by comparing the accelerations measured by two different microscopic quantum systems. This has been achieved using atom interferometry. The current constraints on the UFF between macroscopic and microscopic systems are at the level of $10^{-9}$~\cite{peters:1999qf}. Similar constraints using two different types of atom interferometers have provided constraints at the level of $10^{-7}$~\cite{fray:2004fk}. Furthermore, there is a priori no reason that bodies with different spins react to gravity in a similar way. Therefore, the UFF has also been tested by considering atoms with different spins and the related constraints are at the level of $10^{-7}$~\cite{tarallo:2014fj}. One can wonder if anti-matter falls any differently than SM. Tests of the UFF using anti-hydrogen atoms are currently on-going at CERN with the AEgIS~\cite{aghion:2013nr} and GBAR~\cite{perez:2012ly} experiments.

Even though the existing phenomenological tests of the UFF are quite extensive for SM, the question of the validity of the EEP with DM remains open. Due to the constraints on the UFF for SM, a number of studies have been able to make progress on this question by considering models in which DM is subject to an additional force that is weakly felt by SM. 
These models have a number of testable predictions \cite{Hui2009,Jain2011, Hammami2015}. Simpler models constrain the additional force to DM only. This includes models in which DM has a Yukawa coupling to a light scalar field \cite{1991PhRvL..67.2926F,Gradwohl1992}. It was shown that current galactic observations constrain the violation of the UFF at the level of $10^{-1}$~\cite{2006PhRvD..74h3007K,2006PhRvL..97m1303K}. A similar model of Weakly Interacting Massive Particle has been constrained at the level of $10^{-1}$ with astrophysical observations~\cite{carroll:2009,carroll:2010}. The UFF between two bodies of SM falling in a gravitational field sourced by DM (typically our Galaxy) has also been constrained at the level of $10^{-5}$~\cite{stubbs:1993,schlamminger:2008zr}. Recently, Cosmic Microwave Background observations by the Planck satellite have been used to constrain a similar model at the level of $10^{-6}$~\cite{2015JCAP...10..029B}.

The two other aspects of the EEP are also very well constrained. One way to test the LLI is to search for anisotropies in the speed of the electromagnetic interaction usually parametrized by the $c^2-$formalism~\cite{cliffwill}. Furthermore, an extensive framework called ``Standard Model Extension'' (SME) has been developed to systematically consider Lorentz symmetry violations in all sectors of physics (see for example~\cite{colladay:1997vn,colladay:1998ys,tasson:2014qv}) and an impressive number of the introduced Lorentz violating parameters have been constrained by different experiments (see~\cite{kostelecky:2011ly} for a review of the SME constraints). The LPI is usually tested by redshift experiments~\cite{vessot:1979fk,vessot:1980fk,cacciapuoti:2011ve,delva:2015fk} or by search for space-time variations of the fundamental constants of Nature~\cite{uzan:2011vn,rosenband:2008fk,guena:2012ys,minazzoli:2014xz}. Let us mention that some cosmological consequences of a EEP violations have also been investigated in \cite{hees:2014uq,hees:2015yq}.

In addition to the EEP tests, the second principle of GR (the form of the metric) is thoroughly tested by different observations in the Solar System: deflection of light~\cite{lambert:2009bh}, planetary ephemerides~\cite{pitjeva:2013fk,verma:2014jk,fienga:2015rm,hees:2014jk,hees:2015sf}, Lunar Laser Ranging~\cite{williams:2012zr}, radioscience tracking of spacecraft~\cite{bertotti:2003uq,konopliv:2011dq,hees:2012fk}, etc (for an extensive review, see~\cite{will:2014la}).

In this communication, we address the question of the validity of the EEP for DM. A direct test of the EEP in the Dark sector would consist of a comparison of the acceleration of a body made of DM with the acceleration of a standard test mass. Obviously, this kind of direct test is currently far from being reachable since DM has not been directly detected so far. Nevertheless, we will show that indirect constraints on the EEP in the Dark sector can be reached. These indirect constraints are model dependent.

Phenomenologically, the easiest way to introduce a violation of the EEP in the Dark sector is to couple DM to a metric different from the one to which SM couples. This leads to a large class of theories called bimetric theories of gravitation. A subclass of these theories can be identified when the two metrics are conformally related. The matter part of the action can then be written as
\begin{align}
	S_\textrm{mat}=S_{l}\left[g_{\mu\nu},\Psi_{l}\right]+S_{d}\left[M^2(\Phi)g_{\mu\nu},\Psi_{d}\right]\, ,
\end{align}
where the subscripts $l$ refer to SM and $d$ to DM. The conformal factor $M^2(\Phi)$ depends on a scalar degree of freedom $\Phi$ whose action needs to be specified. A general expression for the gravitational part of the action is given by a generalization of the Brans-Dicke action~\cite{jordan:1949vn,brans:1961fk,brans:2014sc} also sometimes known as the Bergmann-Wagoner framework~\citep{bergmann:1968fk,wagoner:1970fk}
\begin{align} \label{action JF}
	S_\textrm{grav}=\frac{1}{16\pi G_*}\int d^4x \sqrt{-g}\left[h(\Phi)R-\frac{\omega(\Phi)}{\Phi}g^{\mu\nu}\partial_\mu\Phi\partial_\nu\Phi\right] \, ,
\end{align}
where $h$ and $\omega$ are two arbitrary functions of the scalar field $\Phi$. The total action generalizes standard scalar-tensor theories~\cite{jordan:1949vn,brans:1961fk,brans:2014sc,damour:1992ys}. It was first introduced in~\cite{damour:1990fk}. The related cosmological implications have thoroughly been investigated in~\cite{alimi:2007kl,fuzfa:2007mw,alimi:2008zr,alimi:2010kx} where it has been named ``Abnormally Weighting Energy'' (AWE). In particular, it has been shown that the expansion of the Universe can emerge from the DM non-universal coupling~\cite{alimi:2008zr} without any additional potential, cosmological constant or new type of matter.

Note that the EEP violation arising from the action presented above can be highlighted by making use of the Einstein conformal frame. A conformal transformation (see~\cite{alimi:2008zr} for example or the Appendix~\ref{app:einstein}) allows one to write the action as
\begin{align} \label{eq:actionEF}
	S&=\frac{1}{16 \pi G_*}\int d^4x \sqrt{-g_*}\left[R_*-2 g^{\mu\nu}_*\partial_\mu\varphi\partial_\nu\varphi\right] +S_{l}\left[A^2_{l}(\varphi)g^*_{\mu\nu},\Psi_{l}\right]+S_{d}\left[A^2_{d}(\varphi)g^*_{\mu\nu},\Psi_{d}\right]\, ,
\end{align}
where the stars are related to quantities expressed in the Einstein frame, $\varphi$ is a rescaled scalar field and the functions $A_l(\varphi)$ and $A_d(\varphi)$ are the coupling functions between the scalar field and the different types of matter. The action expressed in the Einstein frame is totally equivalent to the original one and is more convenient for some of the calculations~\cite{damour:1992ys,damour:1993uq,damour:1993kx,flanagan:2004fk,hees:2012kx}. In particular, the relative differential acceleration between two test masses\footnote{In order to consider extended bodies, one needs to add a contribution related to the Nordtvedt effect, see~\cite{nordtvedt:1968uq,nordtvedt:1968ys,nordtvedt:1969vn,will:1971zr,will:1989qy,damour:1992ys,hees:2015ty}.}, one made of SM and the other one made of DM, falling in the gravitational field generated by a SM body is given by~\cite{damour:1994fk,damour:1994uq,damour:2010ve,damour:2010zr,minazzoli:2015rz}
\begin{equation}\label{eq:deltaa_a}
	\left(\frac{\Delta a}{a}\right)_{l;l-d}=2\frac{a_l-a_d}{a_l+a_d} = \frac{(\alpha_{l,\infty}-\alpha_{d,\infty})\alpha_{l,\infty}}{1 + \frac{1}{2} \left(\alpha^2_{l,\infty}+\alpha_{d,\infty}\alpha_{l,\infty} \right)} \, ,
\end{equation}
where $a_l$ (resp. $a_d$) is the acceleration of the body made of SM (resp. of DM) and 
\begin{align}
		\alpha_{l,\infty}=\left.\frac{\partial \ln A_{l}(\varphi)}{\partial \varphi}\right|_{\varphi=\varphi_\infty}\, ,\qquad \alpha_{d,\infty}=\left.\frac{\partial \ln A_{d}(\varphi)}{\partial \varphi}\right|_{\varphi=\varphi_\infty}\, .
\end{align}
This shows unambiguously that a difference between the two coupling strengths $\alpha_{l}$ and $\alpha_{d}$ leads to a violation of the UFF and therefore of the EEP. Therefore, any constraints on the parameters $\alpha_{l}$ and $\alpha_{d}$ can be interpreted as an indirect test of the EEP in the Dark sector. We emphasize here that the test proposed in this paper is indirect and depend on the theory considered.

Galactic observations are good candidates to constrain an EEP violation in the Dark sector since DM has a strong impact at these scales. Galactic observations that demonstrate the importance of DM on these scales include galaxy rotation curves, strong gravitational lensing, and velocity dispersion. Galaxy rotation curves are a feature of spiral galaxies as these have a disk in which stars follow circular orbits. However, decomposing the total mass distribution of spiral galaxies into its SM and DM components, a necessary part of testing the EEP on galactic scales, is still an unsolved problem \cite{2010ARA&A..48...87T}. In contrast, strong gravitational lensing combined with velocity dispersion can break down this degeneracy.

Although strong lensing is a very rare phenomenon \cite{2005NewAR..49..387M}, recent advances have greatly improved the number of known galaxy scale lens systems. In this paper, we use combined strong lensing and velocity observations of a sample of  galaxies observed by the Sloan Lens ACS (SLACS) Survey collaboration \cite{2008ApJ...682..964B} to constrain the coupling strengths $\alpha_{l}$ and $\alpha_{d}$. These observations depend on quantities computed in the Jordan frame for SM such as the gravitational potential of the lens and its derivatives as well as the overall geometry of the universe through the angular diameter distances between observer, lens and source. This sample has already been successfully used to constrain the standard post-Newtonian parameter $\gamma_\textrm{PPN}$~\cite{bolton:2006uq,schwab:2010fk} at the level of $5 \times 10^{-2}$ and to constrain a bimetric massive theory of gravity~\cite{enander:2013it}. Our analysis can be considered as an extension of the one from~\cite{bolton:2006uq,schwab:2010fk} which corresponds to the case of a universal coupling characterized by $\alpha_{l}=\alpha_{d}$. 

In Sec.~\ref{sec:model}, we will present the model used in our studies and the main hypothesis underlying our analysis. The derivation of the field equations related to this model and the analytical solution will be sketched in this section while all the calculations are presented in detail in Appendix~\ref{app:WF}. In Sec.~\ref{sec:EinR_and_VelDisp}, we present the observables used in our analysis: the Einstein radius and the velocity dispersion. The dataset used and the details of the statistical analysis performed in this paper are presented in Sec.~\ref{sec:data}. In Sec.~\ref{sec:results}, we present the results from our Bayesian inference and the estimations of the coupling strengths between the different sectors of matter to the scalar field. Finally, our conclusion is presented in Sec.~\ref{sec:conclusion}.

\section{Model}\label{sec:model}
In this section, we will present the main assumptions used in this study and develop the model used in order to describe internal galactic dynamics. First of all, we solve the field equations in the Einstein frame. This frame is more convenient in order to derive the field equations and to solve them in the weak field limit. The variation of the action~(\ref{eq:actionEF}) with respect to the metric and to the scalar field is presented in Appendix~\ref{app:einstein} and leads to the field equations~(\ref{eq:field_einstein}). The invariance of the action under diffeomorphisms also gives a modified conservation equation for the stress-energy tensor. DM and SM are modelled as perfect fluids  and the corresponding Einstein frame stress-energy tensor can thus be expressed as $T^{\mu\nu}_*=\left(\rho_* c^2+p_* \right)u^\mu_* u^\nu_*+p_* g^{\mu\nu}_*$ where $u^\mu_*$ is the Einstein frame 4-velocity. Note that the Einstein frame matter density and pressure can be related to the observable matter density and pressure (in the Jordan frame for SM) by~\cite{damour:1992ys,damour:1993vn,damour:1993kx,damour:1993uq}: 
\begin{align} \label{jord_quants}
\rho_*=A^4_l(\varphi)\rho \, \\
 \mbox{and } p_*=A^4_l(\varphi)p\, .
\end{align}

We will model the galaxies using a static, spherically symmetric model. We use an Einstein frame space-time metric in isotropic coordinates\footnote{We remind the reader that throughout this paper, the stars always refer to quantities expressed in Einstein frame while symbols with no stars refer to quantities expressed in the Jordan frame for SM.}
\begin{equation}\label{eq:metric}
	ds_*^2 = -e^{\nu_*(r_*)}c^2 dt_*^2 + e^{\lambda_*(r_*)} \left( dr_*^2 + r_*^2 d \Omega_*^2 \right) \, .
\end{equation}
This metric differs from the metric expressed in Schwarzschild coordinates used for example in~\cite{damour:1993vn}, which are more suitable to study strong field effects. Isotropic coordinates are more suitable to study observables in weak gravitational fields. The introduction of the metric~(\ref{eq:metric}) in the field equations~(\ref{eq:field_einstein}) leads to the set of equations~(\ref{eq:spherical_field}). Note that these equations can be integrated numerically even for strong fields, although this will not be necessary in this work.

The expression of the matter densities are given by power law profiles in the Jordan frame for SM:
\begin{subequations}\label{eq:densities}
\begin{align}
\rho_{t}(r)&= \rho_{l}(r) + \rho_{d}(r) =  r^{ - \mathbf{\gamma}} {\rho^{(\gamma)}_{0}} \, , \label{eq:DensTot}\\
\rho_{l}(r)&= r^{ - \mathbf{\delta}}  {\rho^{(\mathbf{\delta)}}_{0}} \, , \label{eq:DensLum}
\end{align}
\end{subequations}
where $\rho_{t}(r)$ is the total matter density, $\rho_{l}(r)$ is the SM density and $\rho_d(r)$ is the DM density. A single power profile for the density is valid at the level warranted by the current data \cite{2006ApJ...649..599K}.

We can simplify further our set of equations by using a weak gravitational field expansion. Indeed, the weak gravitational field limit of the field equations can be obtained by using: 
\begin{subequations}
	\begin{align}
		e^{\nu_*} &\approx 1 + \nu_* \, ,\\ 
		e^{\lambda_*} &\approx 1 + \lambda_*  \, .
	\end{align}
\end{subequations}
The weak field limit is justified in the case where the Newtonian potential remains small or equivalently if the galactic compactness parameter $\Xi\sim GM/c^2/R \ll 1$, where $M$ is the total galactic mass included in a volume of radius $R$. At the typical distance corresponding to the observations used in this paper (Einstein radius or radius of stars orbit for velocity dispersion), $\Xi$ is typically of the order of $10^{-7}$, which justifies the weak field approximation used. Moreover, the pressure has to remain small compared to the matter density: $\rho c^2 \gg p$. A careful analysis of the set of equations~(\ref{eq:spherical_field}) shows that another condition needs to be fulfilled in order to reach the weak field limit: $\alpha^2_d \Xi \ll 1$, which for typical galaxies leads to $\alpha_d \ll 10^3$. For very strong coupling strengths of the order of $10^3$, the weak field assumption is not valid. In that case, the scalar field generated by the Dark sector will become very large and will lead to a non-negligible gravitational field. The weak gravitational field limit of the spherical field equations is derived in detail in Appendix~\ref{app:weak_einstein}.

Finally, a last assumption has been made in order to expand the conformal factor
\begin{equation} \label{weak_field_comformal_factor}
	\frac{A_l(\varphi)}{A_{l,\infty}}\approx1+ \alpha_{l,\infty}(\varphi-\varphi_\infty)\, ,
\end{equation}
where $A_{l,\infty}=A_l(\varphi_\infty)$. This expansion is justified as long as $\alpha_{l,\infty}(\varphi-\varphi_\infty)\ll 1$, which is equivalent (see the discussion in Appendix~\ref{app:scalar_field}) to saying that the two parameters $\alpha_{l,\infty}^2$ and $\alpha_{l,\infty}\alpha_{d,\infty}$ remain relatively small (for the galaxies considered in our studies, the approximation remains valid up to $\alpha_{l,\infty}^2\sim 10^3$ and $\alpha_{l,\infty}\alpha_{d,\infty}\sim 10^3$). This assumption can be motivated by a naturalness argument: one expects the coupling strengths between the scalar field and matter to be of the order of unity. Full numerical simulations can be used to study non-linear effects and estimate more properly the domain of validity of the analytical formulas used here. Nevertheless, during our Bayesian inference in Sec.~\ref{sec:results}, we check that we use the formulas in a regime where they are valid.  Indeed, it can be seen in the right of Fig.~\ref{fig:conf_2D}, that the most likely values in the plane $\left(\alpha_{l,\infty}^2,\alpha_{l,\infty}\alpha_{d,\infty}\right)$ are of the order of unity.

The above-mentioned assumptions allow one to derive simple expressions for the metric components, as well as for the scalar field. An analytical solution for the scalar field is found in Appendix~\ref{app:scalar_field}. Although it is easier to  solve the field equations in the Einstein frame, the observables are much more easily derived from the Jordan frame for SM~\cite{damour:1992ys}. Therefore, it is useful to come back to the (SM) Jordan frame metric and introduce the standard gravitational potentials $\Psi$ and $\Phi$ to parametrize this metric as:
\begin{equation}\label{eq:metricMT}
	ds^2=-\left(1+2\frac{\Psi(r)}{c^2}\right)c^2dt^2+\left(1-2\frac{\Phi(r)}{c^2}\right)dl^2 \, .
\end{equation}
These potentials can easily be related to the Einstein frame metric components and to the scalar field as shown in Appendix~\ref{app:weak_jordan}
\begin{subequations}
	\begin{align} \label{pot_weak_field_lim}
		\frac{\Psi(r)}{c^2}&= \alpha_{l,\infty}(\varphi-\varphi_\infty)+\frac{\nu_*}{2}\, ,\\
		\frac{\Phi(r)}{c^2}&= -\alpha_{l,\infty}(\varphi-\varphi_\infty)-\frac{\lambda_*}{2}\, ,
	\end{align}
\end{subequations}
and be used to derive the observables.

Let us summarize the assumptions of our model:
\begin{itemize}
	\item We model the matter components as non-relativistic (pressureless) perfect fluids. 
	\item We use a static, spherically symmetric model and the matter densities are given by expressions~(\ref{eq:densities}).
	\item We use a weak gravitational field expansion justified by the fact that the gravitational potential of a galaxy is $\sim 10^{-7}$ and $\rho c^2>> p$. It is nevertheless important to add that this assumption breaks when $\alpha_{d,\infty}$ becomes of the order of $10^3$.
	\item We use an expansion of the conformal scale factor, which can be done when $\alpha_{l,\infty}^2$ and $\alpha_{l,\infty}\alpha_{d,\infty}$ are not too large (for the galaxies considered lower than $10^3$). This can be justified a priori by a naturalness argument.
\end{itemize}
The strongly coupled regime is not explored in this paper. This regime can lead to a phenomenology that may give rise to spontaneous scalarisation (similar to what has been done in~\cite{damour:1992ys}) and should be explored by making use of numerical simulations.

\section{Einstein radius and velocity dispersion} \label{sec:EinR_and_VelDisp}

\subsection{Einstein radius}\label{sec:einstein}

\begin{figure}[htb]
	\includegraphics[width=0.45\textwidth]{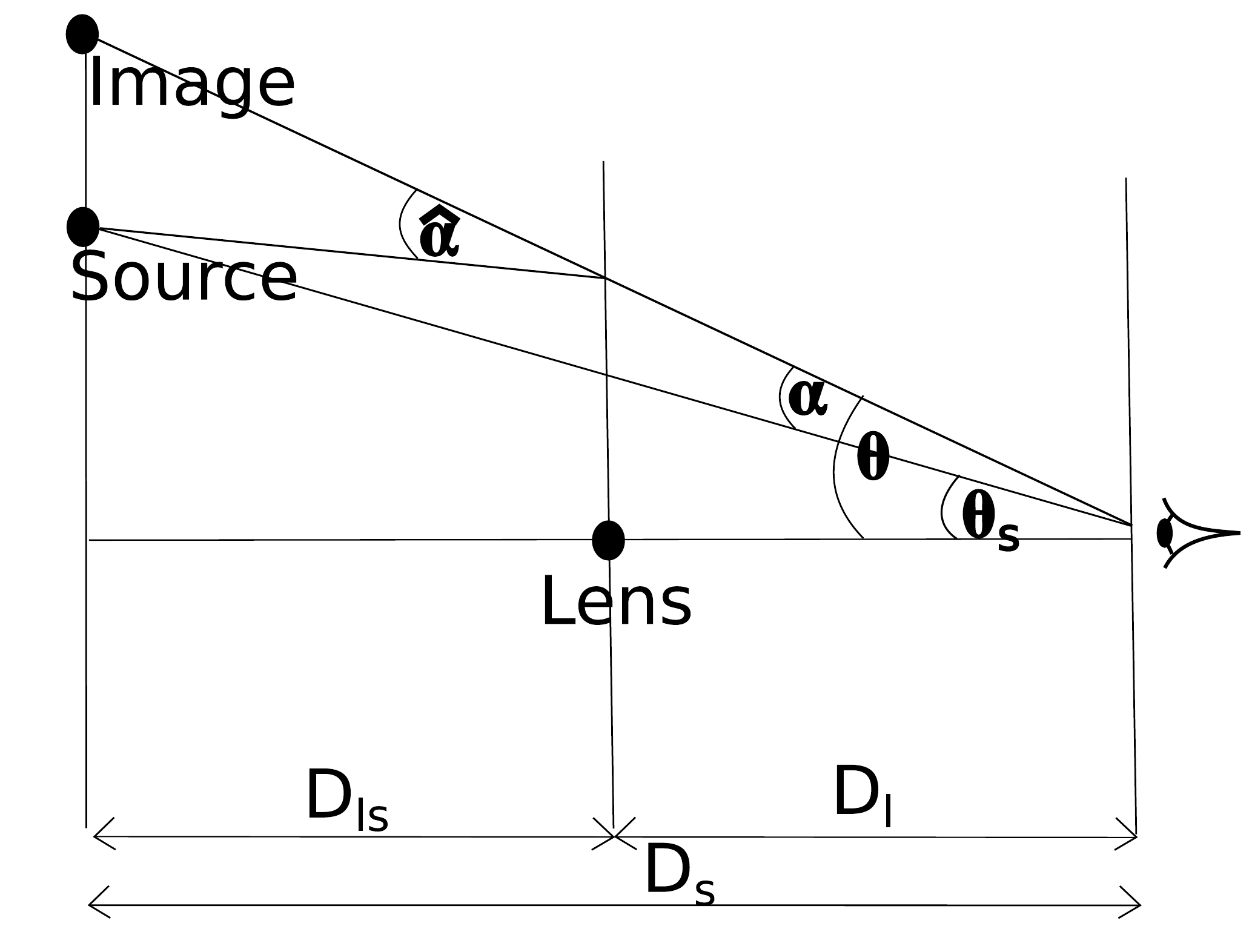}\hfill
\includegraphics[width=0.45\textwidth]{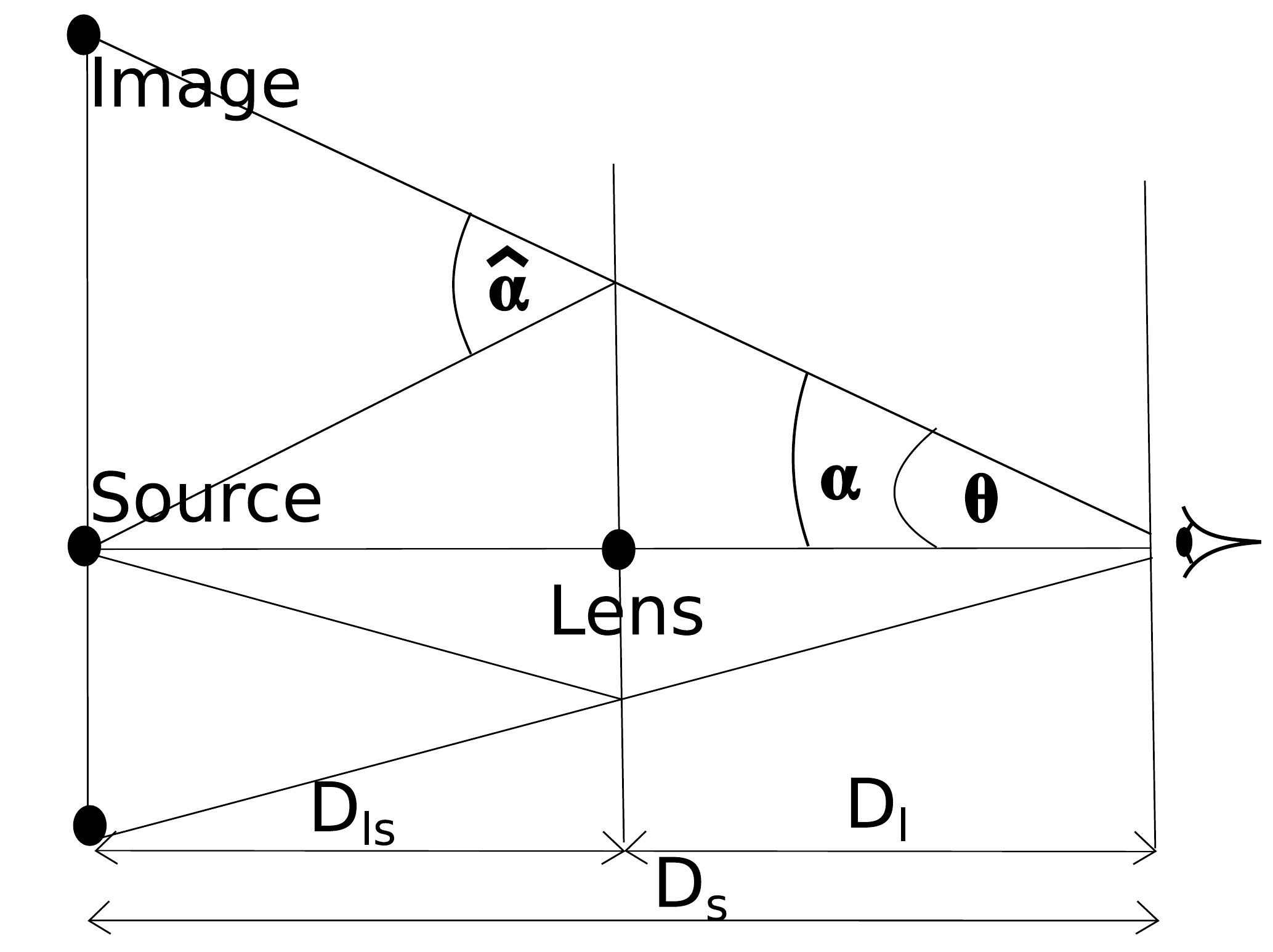}
\caption{The left figure represents a typical illustration of light deflection by a gravitational body. In the special case where the source is aligned with the lens, the image of the source appears as a ring. The right figure illustrates this case.}
\label{fig:geometry}
\end{figure}

An Einstein ring occurs during a lensing event when the source and lens are aligned with respect to the observer. The angular radius of the Einstein ring is called the Einstein radius. A key part to determining the Einstein radius, $\theta_E$, is the deflection angle $\widehat{ \vec{\alpha}}$. For convenience, the optical axis is in the direction of the lens. A real lens is not symmetric, and angles are in general represented by two dimensional vectors. Under the assumption that the deflection angle is small, the angular position of the observed image $\vec{\theta}$, is related to the actual angular position of the source, $\vec{\theta}_S$, by the lens equation \cite{lens_schn}. We introduce the reduced deflection angle $\vec{\alpha}$ as:
\begin{align}
\vec{\alpha} ( \vec{\theta
}) &=  \frac{D_{ls}}{D_s} \widehat{\vec{\alpha}} ( \vec{\theta
}),
\end{align}
where $D_{ls}$ is the angular diameter  distance between the lens and the source and $D_s$ is the angular diameter distance to the source \footnote{The angular diameter distances, $D_l$, $D_s$ and $D_{ls}$ are defined in the observable conformal frame, which is the Jordan frame for SM.}. We then write down the lens equation as
\begin{align} \label{eq::lens_eq_main}
\vec{\theta} &= \vec{\theta}_S +  \vec{\alpha} ( \vec{\theta
})\, .
\end{align}
The geometry and all the quantities introduced are schematically represented in Fig.~\ref{fig:geometry}.
In the case of a circularly symmetric lens, we can disregard the vectorial nature of the quantities in the lens equation. When this is done, we refer instead to the reduced bend angle $\alpha$ instead of its vectorial counterpart, the reduced deflection angle $\vec{\alpha}$. The reduced bend angle is described in detail in Appendix~\ref{def_App} is computed as follows:
\begin{align}
 \alpha\left( \theta \right) &=  \frac{D_{ls}}{D_l D_s}\frac{4G\bar M(\theta)}{\theta}\,
\end{align}
where $D_l$ is the angular diameter distance to the lens, and $\bar{M}$ is the projected mass of a lens within a 2D disk of radius $\theta$. In the weak field limit, one can show that the projected mass can be computed as
\begin{equation}
	\bar M(\theta)=\int_0^{\theta D_l}dx \, 2\pi x \int_{-\infty}^{\infty}dD \frac{A^4_l(\varphi(r))}{A^4_{l,\infty}}\rho_t(r) \, ,
\end{equation}
where the variable $r=\sqrt{D^2+x^2}$ (see Appendix~\ref{def_App}). Using a decomposition of the conformal factor given by Eq.~(\ref{weak_field_comformal_factor}) (see the discussion in Sec.~\ref{sec:model} for the validity condition of such an expansion), one can write all quantities as the sum of a GR part and a contribution coming from the non-zero coupling strengths. The latter will be denoted by a subscript $\varphi$. For example, the expression of the projected mass is given by
\begin{align}
\bar{M}\left( \theta\right) & =  \bar{M}_\textrm{GR}\left( \theta\right) + \bar{M}_{\varphi}\left( \theta\right)\, ,
\end{align}
where the expressions of both terms are given by Eqs.~(\ref{Mass_eq}).
 This results in the following expressions for the reduced deflection angle
\begin{equation}\label{eq:alpha_dec}
\alpha(\theta)=\alpha_\textrm{GR}\left(\theta \right)+\alpha_{\varphi}\left(\theta \right) 	\, ,
 \end{equation}
 with
\begin{subequations} \label{bend_ang}
\begin{align} 
 \alpha_\textrm{GR}\left(\theta \right)= \frac{D_{ls}}{D_lD_s}\frac{4  G\bar{M}_\textrm{GR}\left( \theta\right)}{\theta }   \, ,\\
 \alpha_{\varphi}\left(\theta \right)= \frac{D_{ls}}{D_lD_s}\frac{4  G\bar{M}_{\varphi}\left( \theta\right)}{\theta  } \, .
\end{align}
\end{subequations}
From Eq.~(\ref{eq::lens_eq_main}) it follows that the Einstein radius is a fixed point of the reduced bend angle:
\begin{align} \label{EinR}
\alpha \left( \theta_E\right) &= \theta_E.
\end{align}

Using the decomposition from Eq.~(\ref{eq:alpha_dec}) and a similar one for $\theta_E=\tegr+\theta_{E,\varphi}$, one can solve perturbatively the Einstein radius Eq.~(\ref{EinR}) (see Appendix~\ref{EinR_App}) to find
\begin{subequations}
	\begin{align}
		\tegr&= \left(\frac{D_{ls}}{D_sD_l}4  G  \bar{M}_\textrm{GR}\left( \theta_{E, GR}\right) \right)^{\frac{1}{2}} \, ,\\
	\theta_{E,\varphi} &= \frac{\alpha_\varphi(\tegr)}{1-\left.\frac{\partial \alpha_\textrm{GR}}{\partial \theta}\right|_{\theta=\tegr}} \, .
	\end{align}	
\end{subequations}
The detailed expressions of these two quantities are given by Eqs.~(\ref{defl}).
 
\subsection{Velocity dispersion}

The line-of-sight stellar velocity dispersion is a measure of how much stellar velocities vary from the mean line-of-sight stellar radial velocity \cite{2008gady.book.....B}. In spherical symmetry, it is given by the expression \cite{1980MNRAS.190..873B}:
\begin{align}\label{eq:sigma}
\sigma_r^2 \left( r \right) &=\frac{1}{\rho_{l}( r ) r^{2 \beta}} \int^\infty_r  \rho_{l} ( x ) x^{2 \beta} \dot \Psi(x) dx \, ,
\end{align}
where $\rho_{l}(r)$ is the SM density given by Eq.~(\ref{eq:DensLum}), $\Psi$ is the gravitational potential appearing in the time-time component of the space-time metric~(\ref{eq:metricMT}), and  $\beta = 1 - \frac{\sigma^2_t}{\sigma_r^2}$ is the velocity anisotropy parameter in which $\sigma_t$ and $\sigma_r$  are the tangential and radial components of the velocity dispersion. The dot represents the derivative with respect to $r$, the radial coordinate in the Jordan frame for SM. 

To obtain the last expression, it has been implicitly assumed that the motion of stars is governed by the gradient of the Jordan frame Newtonian potential (i.e. the temporal part of the Jordan frame metric, see Eq.~(\ref{eq:eq_mot_vel_disp})). This means that we are neglecting potential strong field deviations that may arise from a Nordtvedt effect~\cite{nordtvedt:1968uq,nordtvedt:1968ys,nordtvedt:1969vn,will:1971zr,will:1989qy,damour:1992ys,hees:2015ty}. In the theory considered in this paper, such an effect will be parametrized by a difference between active and passive mass of the form~\cite{will:2014la,damour:1992ys,hees:2015ty}\footnote{This difference is characteristic of a violation of the Strong Equivalence Principle and therefore arises even if $A_l=A_d$, or in other words even when the EEP is not violated.}
\begin{equation}
	\frac{m_P}{m_A}\approx1+ \frac{2 \alpha_{l,\infty}^2}{1+\alpha_{l,\infty}^2}\frac{\left|\Omega_g\right|}{m_Ac^2}\, ,
\end{equation}
where $\Omega_g$ is the gravitational self-energy of the body. In the case of stars, the gravitational self-energy is typically of the order of $\left|\Omega_g\right|/m_Ac^2\sim 10^{-6}$. The Nordtvedt effect will therefore produce a relative modification of the estimated $\sigma^2_r$ of the order of $10^{-6}\alpha^2_{l,\infty}$. Since the relative accuracy of the velocity dispersion measurements is of the order of 5-10 \%~\cite{2008ApJ...682..964B,enander:2013it}, it is safe to neglect the Nordtvedt effect as long as $\alpha_{l,\infty}<10$. Note that as already mentioned previously in this paper, we do not explore the region of strong coupling strength and we can safely neglect this effect.

In practice what is observed is the luminosity averaged over the line-of-sight, weighted over the spectroscopic aperture, $\sigma_{\star}$, which is given by the expression \cite{2008gady.book.....B}
\begin{align}
\sigma^2_{\star} &= \frac{\int_0^\infty dR \, R \, w(R) \int_{-\infty}^\infty dz \rho_{l}(r) \left( 1 - \beta  \frac{R^2}{r^2 }\right) \sigma_r^2(r)}{\int_0^\infty dR \, R\, w(R)\int_{-\infty}^\infty dz \rho_{l} (r)} \, ,\label{eq:sigmastar}
\end{align}
where $r^2 = R^2 + z^2$, $w\left( R \right) = e^{-R^2/2 \bar{\sigma}_{atm}^2}$ is the aperture weighting function depending on the atmospheric seeing $\bar{\sigma}_{atm}$  (see~\cite{2006ApJ...638..703B,2008ApJ...682..964B}) and $\sigma_r(r)$ is the radial velocity dispersion of the SM given by Eq.~(\ref{eq:sigma}).

Similar to what has been done in the computation of the Einstein radius in Sec.~\ref{sec:einstein}, we decompose all the quantities as the sum of a GR contribution and a contribution from the scalar field. The Eq.~(\ref{pot_weak_field_lim}) allows one to write
\begin{align}
\Psi = \Psi_\textrm{GR} + \Psi_{\varphi} \, ,
\end{align}
which can be used in Eqs.~(\ref{eq:sigma}) and (\ref{eq:sigmastar}) to compute the velocity dispersion averaged over the line-of-sight luminosity as
\begin{align}
\sigma^2_{\star} &= \sigma^2_{\star, GR} + \sigma^2_{\star, \varphi}\, .
\end{align}
The computation is done explicitly in Appendix \ref{app:vel_disp} and the full expressions of the two terms appearing in the last equation are given by Eqs.~(\ref{eq:vel_disp_phi}).

\section{Procedure to analyse data} \label{sec:data}

Approximately 200 cases of strong gravitational lensing by galaxies are known to date \cite{2010ARA&A..48...87T}. In the last 15 years, a large number of these were discovered in four surveys: the Cosmic Lens All-Sky Survey (CLASS) \cite{2003MNRAS.341....1M}, the Sloan Digital Sky Survey (SDSS) Quasar Lens Search (SQLS) \cite{2008AJ....135..496I}, the Sloan Lens ACS Survey SLACS \cite{2007ApJ...667..176G} and the Hubble Space Telescope COSMOS survey \cite{2008ApJS..176...19F}. Strong lenses can be classified into two categories: galaxy-galaxy lenses and galaxy-quasi-stellar object lenses. The former are suitable for studying the gravitational potential of the lens itself, as the emission is not overwhelmed by the source which is the case in the rarer galaxy-quasi-stellar lenses \cite{2005NewAR..49..387M}. Despite the increasing wealth of observational data, most known galaxy strong lens systems are found at a redshift $z \lesssim 0.4, $ and the source or deflector redshift are often missing. Future surveys including the Dark Energy Survey~\cite{DES}, Euclid~\cite{amendola:2013ab}, the Large Synoptic Survey Telescope \cite{2015ApJ...811...20C} and the Square Kilometre Array \cite{2015aska.confE..84M} are expected to increase the number of known lenses by a factor of $10^3$.

\subsection{Dataset}

The data used in our analysis is a sub-sample of 53 galaxies presented in Table 4 of \cite{2008ApJ...682..964B}\footnote{A compilation of the data conveniently containing the redshifts, velocity dispersions, Einstein radii, and SM power law index $\delta$ of the galaxies used in this study can be found at \url{http://www.fysik.su.se/~edvard/slacsl.html}, see~\cite{enander:2013it}.}. Complementary studies on the same data have examined the validity of GR by constraining the post-Newtonian parameter $\gamma_\textrm{PPN}$ \cite{bolton:2006uq,schwab:2010fk} and Hassan-Rosen bimetric gravity \cite{enander:2013it}.

The complete dataset consists of the line-of-sight luminosity velocity dispersion $\sigma_{\star}$ and its standard error $\varepsilon_g$, the Einstein radius $\theta_*$, the SM profile parameter $\delta$, the redshift to the source $z_s$, the redshift to the lens, $z_l$, the luminosity of each galaxy in the visual band $\lambda = 555 \ \mbox{nm}$, $L_{V555}$, and the effective radius of each galaxy $R_g$. 

The fit cannot be done without three additional parameters, the Hubble rate $H_0$ and the deceleration parameter $q_0$, as well as the aperture parameter $\bar{\sigma}_{atm}$. We will set $H_0$ to exactly 67.8 km/s/Mpc, and further assume a flat universe with $\Omega_{\Lambda} = 0.692$ to determine $q_0$. These values correspond to the best fits values obtained by Planck \cite{2015arXiv150201589P}. In Sec.~\ref{sec:influence_cosmo}, we show that our analysis is robust to a change in these two parameters. The parameter $\bar{\sigma} = 1.6''$, is chosen to be consistent with the range determined by \cite{2008ApJ...682..964B}.

\subsection{Velocity dispersion computation}\label{sec:velocity_disp}
In this section, we will present briefly the procedure used to compute a velocity dispersion. The calculations related to this section are presented in Sec.~\ref{sec:EinR_and_VelDisp} and derived in detail in Appendix~\ref{app:lensing},~\ref{app:vel_disp} and~\ref{DensPar_App}. First, let us list the parameters considered in the Bayesian inference and the parameters directly inferred from observations. The observed parameters are (for each galaxy): the Einstein radius $\theta_*$, the exponent for the SM density distribution $\delta$, the redshifts of the source $z_l$ and the lense $z_s$, the luminosity $L_{V555}$ included in an effective radius $R_g$. The parameters considered in the Bayesian inference are: the two coupling strengths $\alpha_{l,\infty}$, $\alpha_{d,\infty}$, the power-law exponent  $\gamma$ that parametrizes the total matter profile (see Eq.~(\ref{eq:DensTot})), the velocity dispersion anisotropy parameter $\beta$ and the mass-to-light ratio $\Upsilon$ between the SM mass of the galaxy ($M_*$) and the $L_{V555}$. The variable of interest is the velocity dispersion $\sigma_\star$. The procedure to compute $\sigma_\star$ consists in the following steps:
\begin{itemize}
 \item 
The Hubble parameter $H_0$ and the cosmic deceleration parameter $q_0$ are used to transform the redshifts $z_l$ and $z_s$ into angular diameter distances $D_l$, $D_s$ and $D_{ls}$. 
 
 \item The value of $M_*=\Upsilon L_{V555}$ and of $R_g$ are used with Eq.~(\ref{eq:rho_d_GR}) to evaluate the GR part of the SM density parameters $\rho_{0,\textrm{GR}}^{(\delta)}$.
 
 \item Using the observed value of the Einstein radius $\theta_*$, one can evaluate the GR part of the total density parameter $\rho_{0,\textrm{GR}}^{(\gamma)}$  with Eq.~(\ref{eq:rho_g_GR}).
 
 \item 
The computed values of $\rho_{0,\textrm{GR}}^{(\delta)}$ and $\rho_{0,\textrm{GR}}^{(\gamma)}$ along with Eqs.~(\ref{eq:ks}) are used to evaluate the constants $k_\delta$ and $k_\gamma.$
 
 \item It is then possible to evaluate the first order correction to the density parameters $\rho_{0}^{(\delta)}$ and $\rho_{0}^{(\gamma)}$. The  correction $\rho_{0,\varphi}^{(\delta)}$ is determined from Eq.~(\ref{eq:rho_d_phi}) where $\bar M_\textrm{lum}$ is given by (\ref{eq:mlumphi}). The correction $\rho_{0,\varphi}^{(\gamma)}$ is given by Eq.~(\ref{eq:rho_g_phi}).
 
 \item Lastly, using Eqs.~(\ref{eq:vel_disp_gr}) and (\ref{eq:vel_disp_phi}) one can evaluate $\sigma_\star.$ 
 
\end{itemize}
\subsection{Bayesian inference}
In this paper, we perform a Bayesian inference to constrain the coupling strengths $\alpha_{l,\infty}$ and $\alpha_{d,\infty}$ that characterize a possible violation of the EEP in the Dark sector. Three additional parameters need to be considered in this inference: $\gamma$, $\beta$ and $\Upsilon$. Since we are not interested in those, we will marginalize over these at the end of the process.

The observations of the 53 galaxies (i.e. the velocity dispersions) are assumed to be independent and their distribution to follow an inverse Gamma distribution. This distribution is justified because the velocity dispersions are required to be positive and because this distribution is the conjugate distribution on the variance parameter of a Gaussian \cite{gelman2013bayesian}. The probability density function (pdf) describing the likelihood is given by
\begin{equation}
 L(\sigma_{\star,g}| \alpha_{l,\infty},\alpha_{d,\infty},\gamma,\beta,\Upsilon)=\prod_{g=1}^{N_g}\frac{\beta_g^{\alpha_g}}{\Gamma(\alpha_g)}\sigma_{\textrm{sim,g}}^{-1-\alpha_g}\exp\left(-\frac{\beta_g}{\sigma_{\textrm{sim,g}}}\right) \, ,
\end{equation}
with $N_g$ the number of galaxies, $\Gamma(x)$ the Gamma function, $\sigma_{\textrm{sim,g}}$ are the simulated values of the velocity dispersions that depends on the parameters considered. The procedure to compute the prediction of the velocity dispersion ~$\sigma_{\textrm{sim,g}}=\sigma_\star(\alpha_{l,\infty},\alpha_{d,\infty},\gamma,\beta,\Upsilon;\allowbreak L_{V555,g},\theta_{*,g},\delta_g )$ is fully described in Sec.~\ref{sec:velocity_disp}. Moreover, $\alpha_g, \beta_g$ are the parameters characterizing each individual likelihood. They can be related to the mean $\sigma_{\star,g}$ and variance $\varepsilon_g^2$ of each observations through \cite{klugman2012loss}
\begin{align}
\sigma_{\star,g}&=\frac{\beta_g}{\alpha_g-1} \, , \qquad \varepsilon_g^2=\frac{\beta_g^2}{(\alpha_g-1)^2(\alpha_g-2)} \, .
\end{align}

The posterior pdf is given by
\begin{equation}
 p( \alpha_{l,\infty},\alpha_{d,\infty},\gamma,\beta,\Upsilon| \sigma_{\star,g})=\mathcal C\, L(\sigma_{\star,g}| \alpha_{l,\infty},\alpha_{d,\infty},\gamma,\beta,\Upsilon) \pi(\alpha_{l,\infty},\alpha_{d,\infty},\gamma,\beta,\Upsilon) \, ,
\end{equation}
where $\mathcal C$ is a normalizing constant and $\pi(\alpha_{l,\infty},\alpha_{d,\infty},\gamma,\beta,\Upsilon)$ is the prior pdf on the parameters. We assume the parameters to be a priori independent so that 
\begin{equation}
\pi(\alpha_{l,\infty},\alpha_{d,\infty},\gamma,\beta,\Upsilon)=\pi_{\alpha_l,\alpha_d}(\alpha_{l,\infty},\alpha_{d,\infty}) \pi_\gamma(\gamma) \pi_\beta(\beta) \pi_\Upsilon(\Upsilon)\, ,
\end{equation}
with $\pi_{\alpha_l,\alpha_d}(\alpha_{l,\infty},\alpha_{d,\infty})$ the 2D prior pdf on the coupling strengths. In the general case, we assume the prior on the two coupling strengths  to be independent $\pi_{\alpha_l,\alpha_d}(\alpha_{l,\infty},\alpha_{d,\infty})=\pi_{\alpha_l}(\alpha_{l,\infty})\pi_{\alpha_d}(\alpha_{d,\infty})$. Nevertheless,  let us mention that in Sec.~\ref{sec:BD} and~\ref{sec:AWE}, we enforce the coupling strengths to be equal or of opposite signs $\alpha_{l,\infty}=\pm \alpha_{d,\infty}$. In these cases, the number of independent parameters considered in the analysis is reduced by one. This corresponds as taking a non-independent prior $\pi_{\alpha_l,\alpha_d}(\alpha_{l,\infty},\alpha_{d,\infty})=\pi_{\alpha_l}(\alpha_{l,\infty})\delta(\alpha_{l,\infty}\mp\alpha_{d,\infty})$. We will discuss the form of the individual prior pdf's used in this study in Sec.~\ref{sec:priors}. 

Finally, the marginal pdf of the two parameters of interest $\alpha_{l,\infty}$ and $\alpha_{d,\infty}$ is obtained by
\begin{equation}
 p( \alpha_{l,\infty},\alpha_{d,\infty} | \sigma_{\star,g})= \int d\Upsilon\int d\gamma \int d\beta \, p( \alpha_{l,\infty},\alpha_{d,\infty},\gamma,\beta,\Upsilon| \sigma_{\star,g}) \, .
\end{equation}

The MCMC algorithm used is a standard Metropolis-Hastings algorithm based on a simplified version of the software developed in \cite{raey}. We run the Metropolis-Hasting sampler until $10^7$ samples have been generated. The convergence of the Monte Carlo is ascertained by monitoring the credible region. Finally, to diminish the effect of the starting configuration, we discard the first $10^4$ samples.

\subsubsection{Prior pdf's used}\label{sec:priors}
For practical considerations, we use results from previous studies to guide us in choosing reasonable prior pdf's. Different priors have been used in this analysis.
\begin{itemize}
 \item Typical values for the anisotropy velocity dispersion $\beta$ for a sample of nearby galaxies are mentioned in \cite{schwab:2010fk}. The values for $\beta$ range from $-\infty$ to 1. In this paper, we use a beta prime distribution for $1-\beta$. This distribution is justified because of the range of value for $\beta$ but also because using an inverse Gamma distribution for $\sigma_t^2$ and $\sigma_r^2$ (see the justification of the likelihood above) leads to a beta prime distribution on $1-\beta$. The parameters of this distribution are determined so that the mean value for $\beta$ is $< \beta>=0.18$ and the corresponding standard deviation is $\sigma_\beta=0.13$. These values are the same as the ones used in~\citep{bolton:2006uq,schwab:2010fk,enander:2013it} and corresponds to the distribution of mass-dynamical properties of the well-studied sample of nearby elliptical galaxies from \cite{2001AJ....121.1936G}. Note that we checked the robustness of our results by using a uniform prior pdf  and a normal prior pdf. The results are very robust with respect to these changes.
 
 \item The total density profile is known to be close to an isothermal profile. We use normal prior pdf  characterized by mean $< \gamma>=2$ and a standard deviation $\sigma_\gamma=0.08$~\cite{schwab:2010fk}. We checked the robustness of our results by using a uniform prior distribution as well.

 \item The authors in \cite{2007ApJ...667..176G} found that on average $\Upsilon = 4.48 \pm 0.46 h \frac{M_\odot}{L_\odot}$ (at $z = 0.2$) out to 100 effective radii for a sub-sample of the data considered in this study. The prior pdf used for the $\Upsilon$ parameter is  a flat prior pdf  between 1 and 10 to preserve the order of magnitude of the mass-to-light ratio.
 
 \item No a priori knowledge is found on $\alpha_{d,\infty}$. Therefore, we use a non-informative scale invariant Jeffreys prior pdf for this parameter (see the discussion in chapter 3 of~\cite{gregory:2010qv}). This prior is scale invariant and is also invariant under reparametrization.
 
 \item Two different prior pdf's have been considered for the parameter $\alpha_{l,\infty}$. In the first case where only galactic data are used, we assume no prior knowledge on this parameter and use a non-informative scale invariant Jeffreys prior pdf (see the discussion in chapter 3 of~\cite{gregory:2010qv}). In the second case, we use additional information from Solar System observations. More precisely, it is known that the action~(\ref{eq:actionEF}) leads to an expression of the parametrized post-Newtonian (PPN) parameter $\gppn$ given by  \cite{damour:1992ys}
 \begin{equation}\label{eq:gppn}
   \gppn -1 = -2 \frac{\alpha_{l, \infty}^2}{1 + \alpha_{l, \infty}^2}\, ,
 \end{equation}
where we implicitly assume that DM can be neglected in the Solar System. Moreover, this parameter has been estimated by the Cassini spacecraft as~\cite{bertotti:2003uq}
  \begin{align}
    \gppn -1 =(2.1\pm 2.3)\times 10^{-5} \, .
  \end{align}
  The second prior used in this analysis comes from this constraint. We transform the Cassini estimation into a prior pdf for the $\alpha_{l,\infty}$ parameter.  The two different priors considered for $\alpha_{l,\infty}$ will lead to different results as we discuss in the Sec.~\ref{sec:results}.
\end{itemize}

\section{Results} \label{sec:results}

In this section, we will present the main results of our Bayesian inference. All the posterior pdf's presented on $\alpha_{l,\infty}$ and $\alpha_{d,\infty}$ are marginalized over the other parameters $\gamma$, $\beta$ and $\Upsilon$. As mentioned in Sec.~\ref{sec:priors}, different types of prior pdf's have been considered for $\gamma$ and $\beta$. In the following, all the results presented have been obtained with a normal prior for $\gamma$, a beta prime distribution on $1 - \beta$. Nevertheless, as mentioned in Sec.~\ref{sec:priors}, our results are robust with respect to a change of prior on these two parameters.

\subsection{General case with two independent couplings}

We first consider the general case where the two coupling strengths $\alpha_{l,\infty}$ and $\alpha_{d,\infty}$ are independent. It is interesting, although not unexpected, to notice that the results are completely symmetric under a change of sign of the two coupling strengths: $\alpha_{l,\infty}\rightarrow - \alpha_{l,\infty}$ and $\alpha_{d,\infty}\rightarrow -\alpha_{d,\infty}$. This comes from the fact that the observables derived in the Appendix~\ref{app:lensing} and~\ref{app:vel_disp} only depend on the two combinations: $\alpha^2_{l,\infty}$ and $\alpha_{l,\infty}\alpha_{d,\infty}$ and is a general feature of this type of tensor-scalar theory~\citep{damour:1992ys}.

\subsubsection{Analysis using only galactic observations}

In this section we use only galactic observations to constrain the two coupling strengths. This means that we have used a non-informative Jeffreys prior on both coupling strengths (see the discussion in Sec.~\ref{sec:priors}).

\begin{figure}[htb]
	\includegraphics[width=0.45\textwidth]{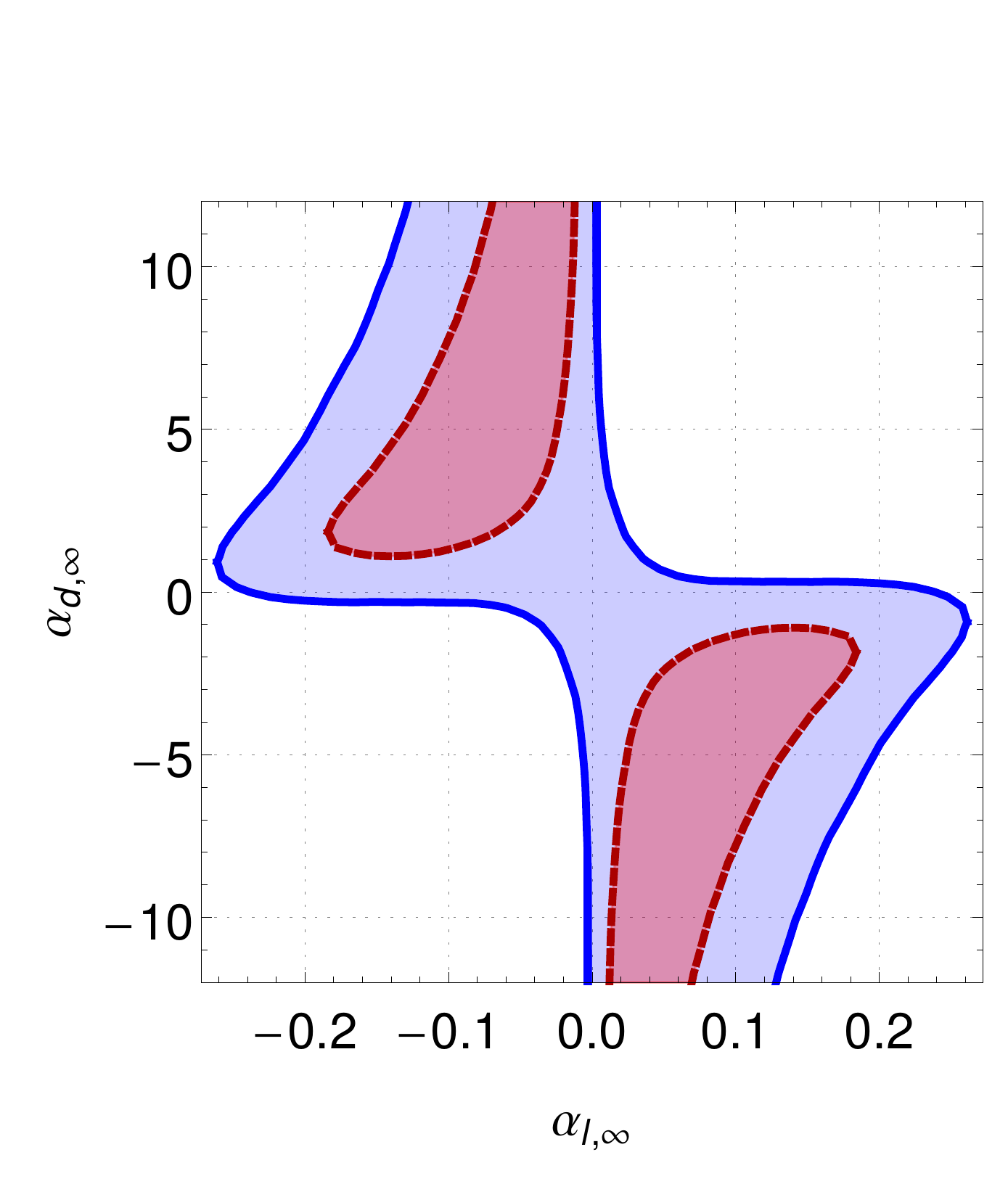}\hfill
\includegraphics[width=0.45\textwidth]{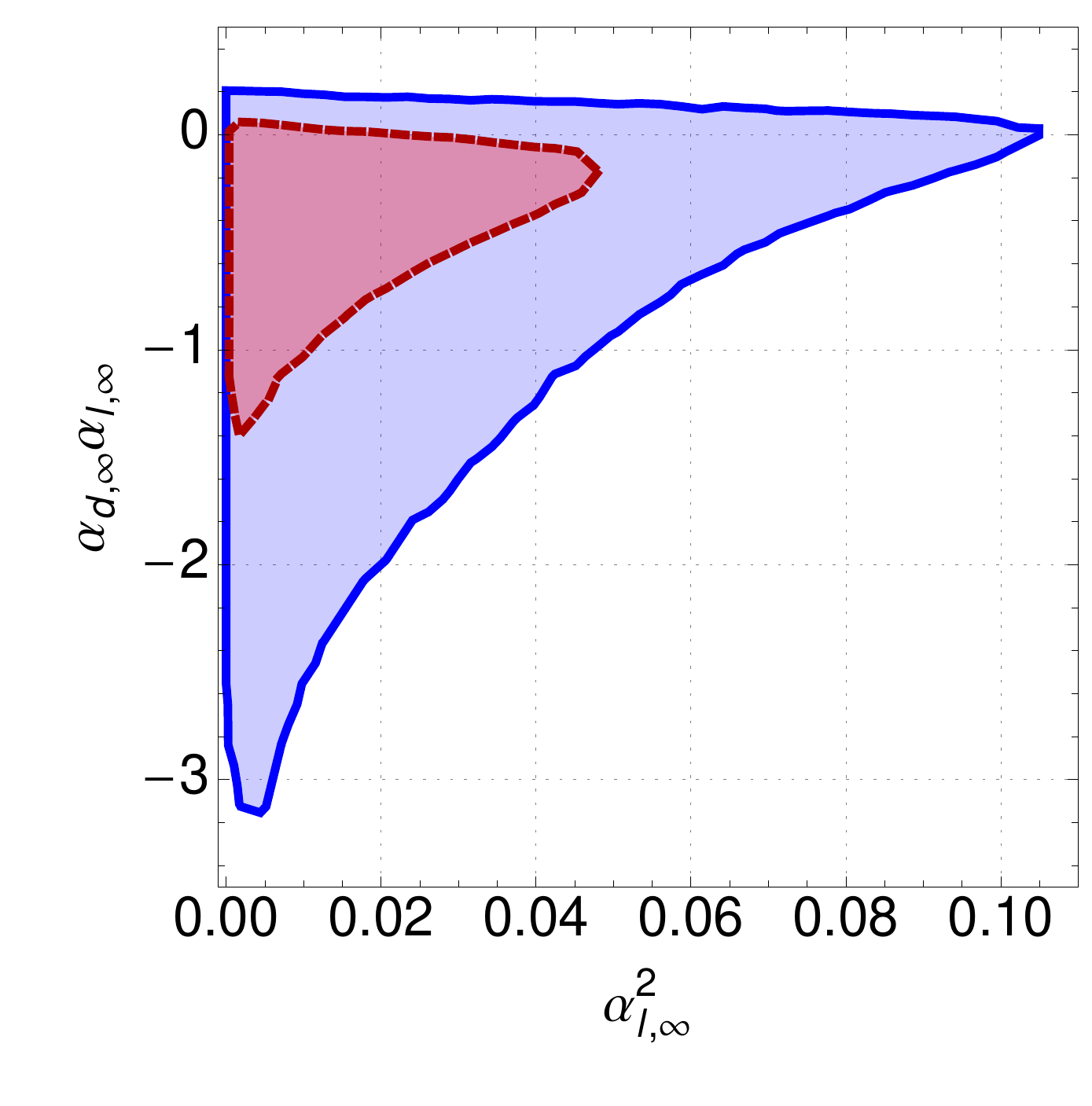}
\caption{Representation of the 68 \% (red, dashed lines) and 95 \% (blue, continuous line) credible regions obtained using only galactic observations.}
\label{fig:conf_2D}
\end{figure}

\begin{figure}[htb]
\centering
\includegraphics[scale=0.4]{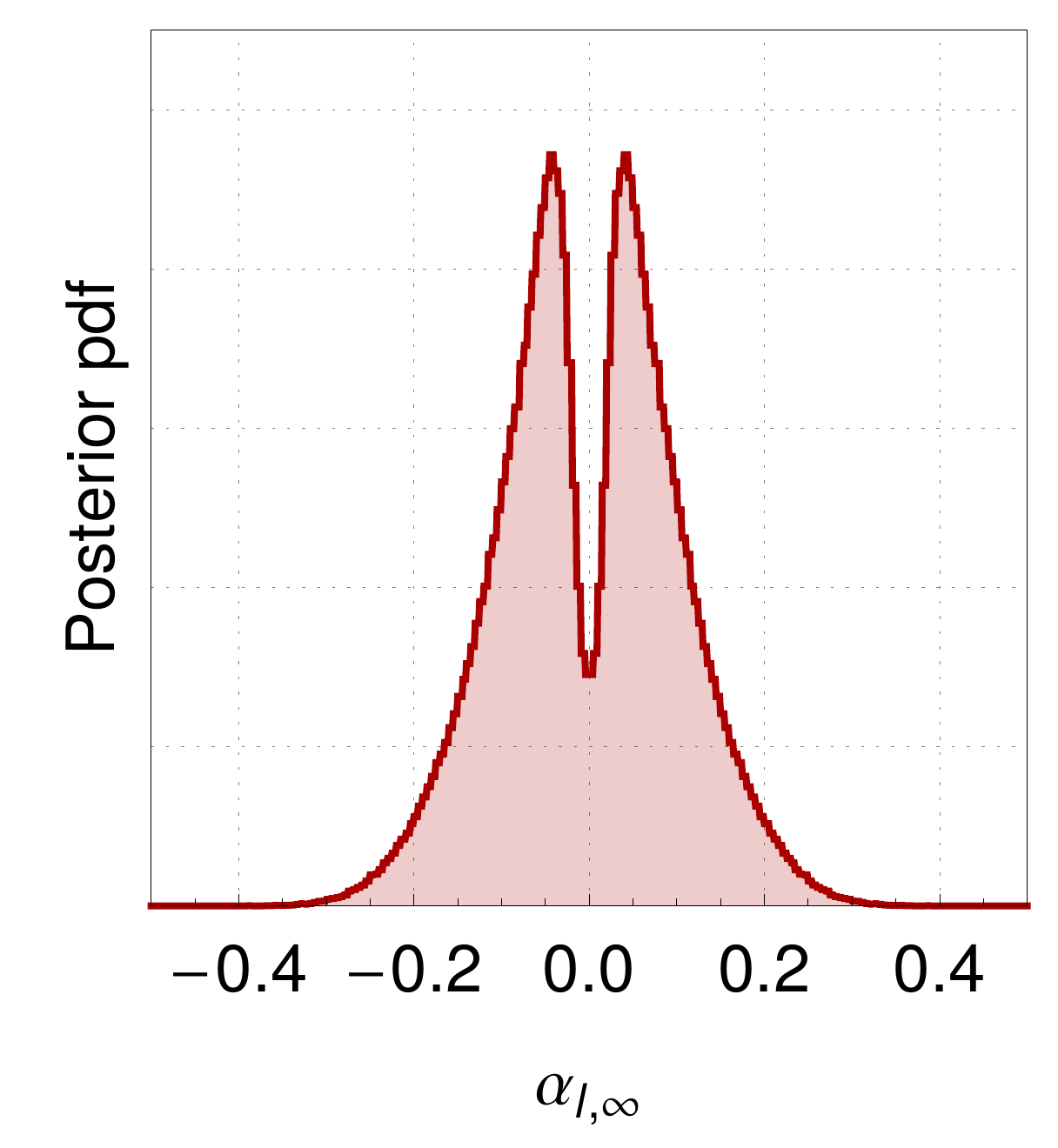}\hspace{0.12\linewidth}
\includegraphics[scale=0.4]{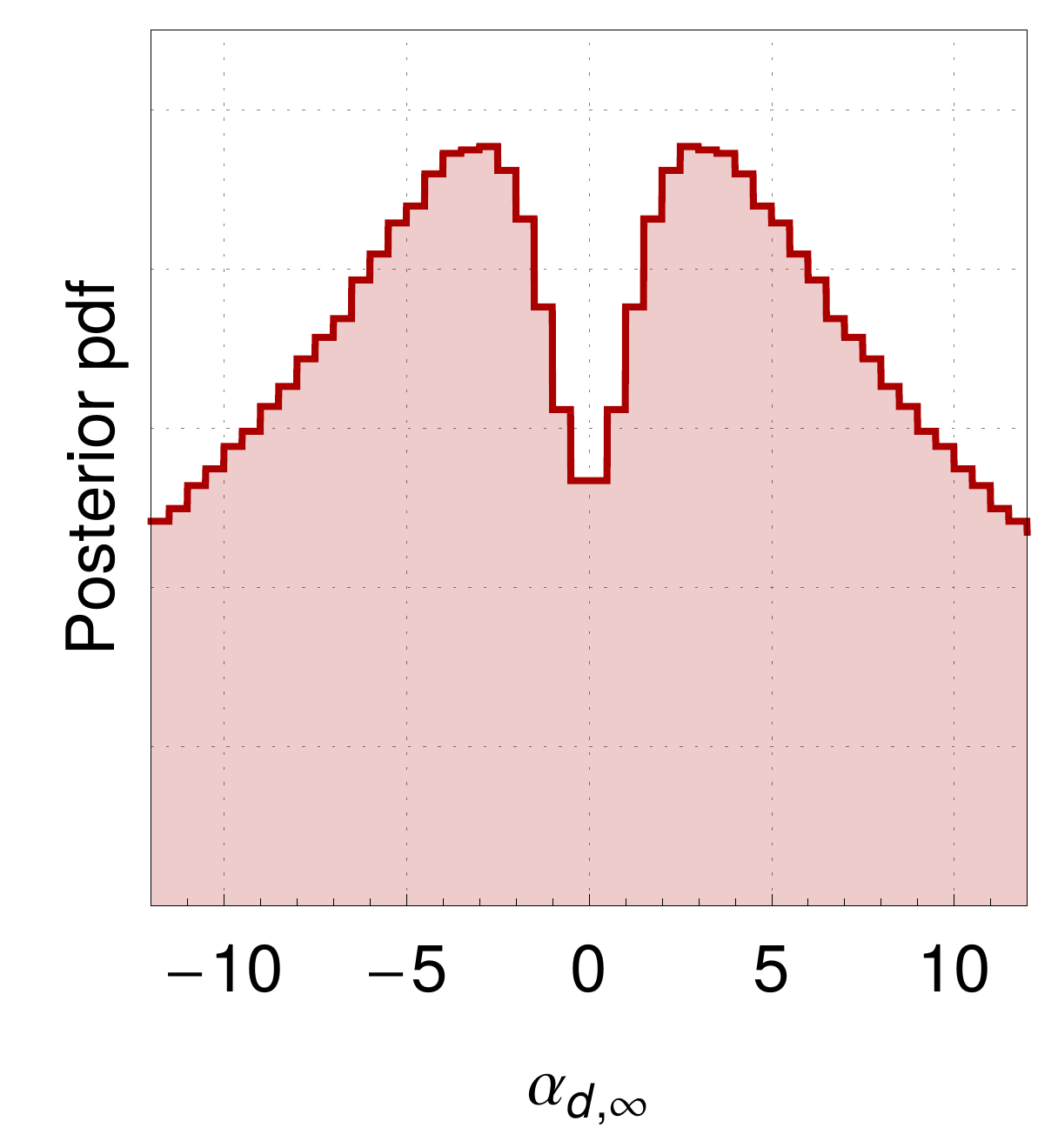}
\caption{Representation of the 1D marginalized posterior pdf inferred by our Bayesian analysis on the coupling strengths $\alpha_{l,\infty}$ and $\alpha_{d,\infty}$ obtained using only galactic observations. }
\label{fig:hist_alpha}
\end{figure}

The left part of Fig.~\ref{fig:conf_2D} shows the marginal 68 \% and 95 \% credible regions obtained by our Bayesian analysis. The symmetry related to the change of sign of the coupling strengths is obvious. The regions of parameters favored by the data are characterized by coupling strengths of opposite signs, $\alpha_{l,\infty}\alpha_{d,\infty}<0$. These have a ``hyperbolic-like'' shape. The data does not constrain the extent of $\alpha_{d,\infty}$. This can be clearly seen in Fig.~\ref{fig:hist_alpha} where the 1D histograms of the coupling strengths obtained by the MCMC are presented. The marginalized posterior pdf on $\alpha_{l,\infty}$ has a bimodal shape. Similarly, the marginalized posterior pdf on $\alpha_{d,\infty}$ presents a bimodal shape and extends to very large values. Let us note again that the approximations used in our modelling break when $\alpha_{d,\infty}\sim 10^3$ (see Sec.~\ref{sec:model}).

\begin{figure}[htb]
\includegraphics[width=0.32\textwidth]{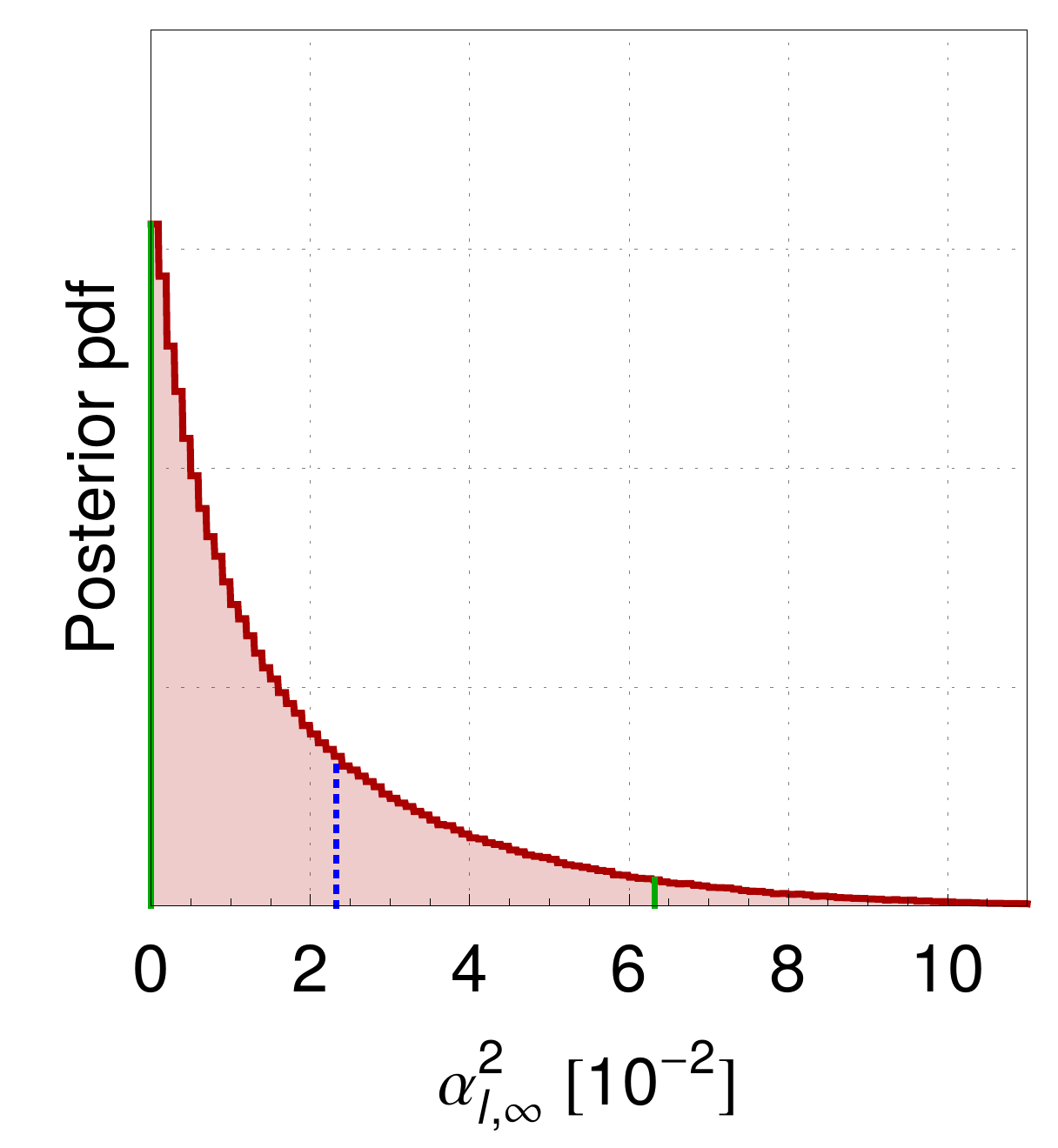}\hfill
\includegraphics[width=0.32\textwidth]{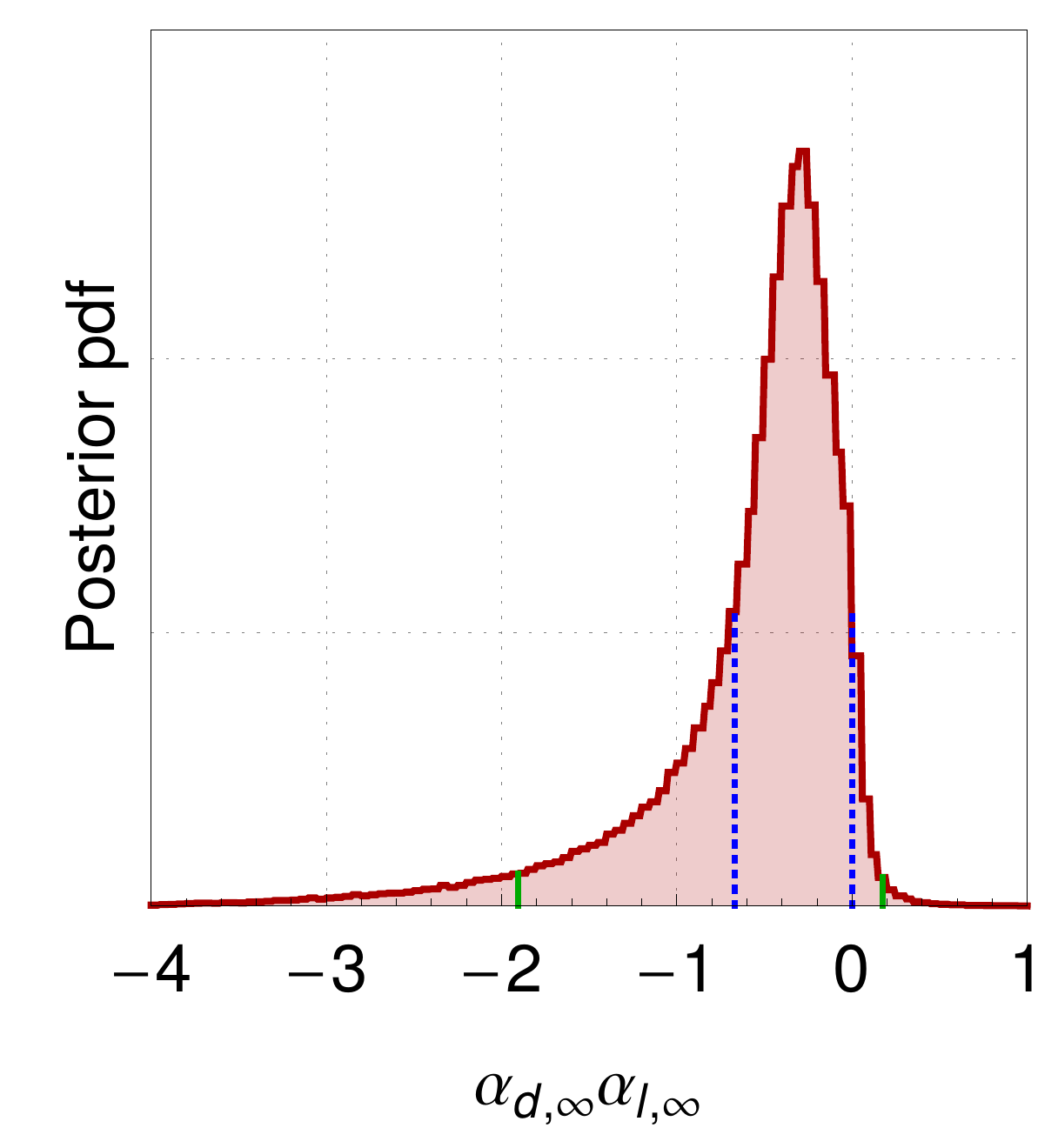}\hfill
\includegraphics[width=0.32\textwidth]{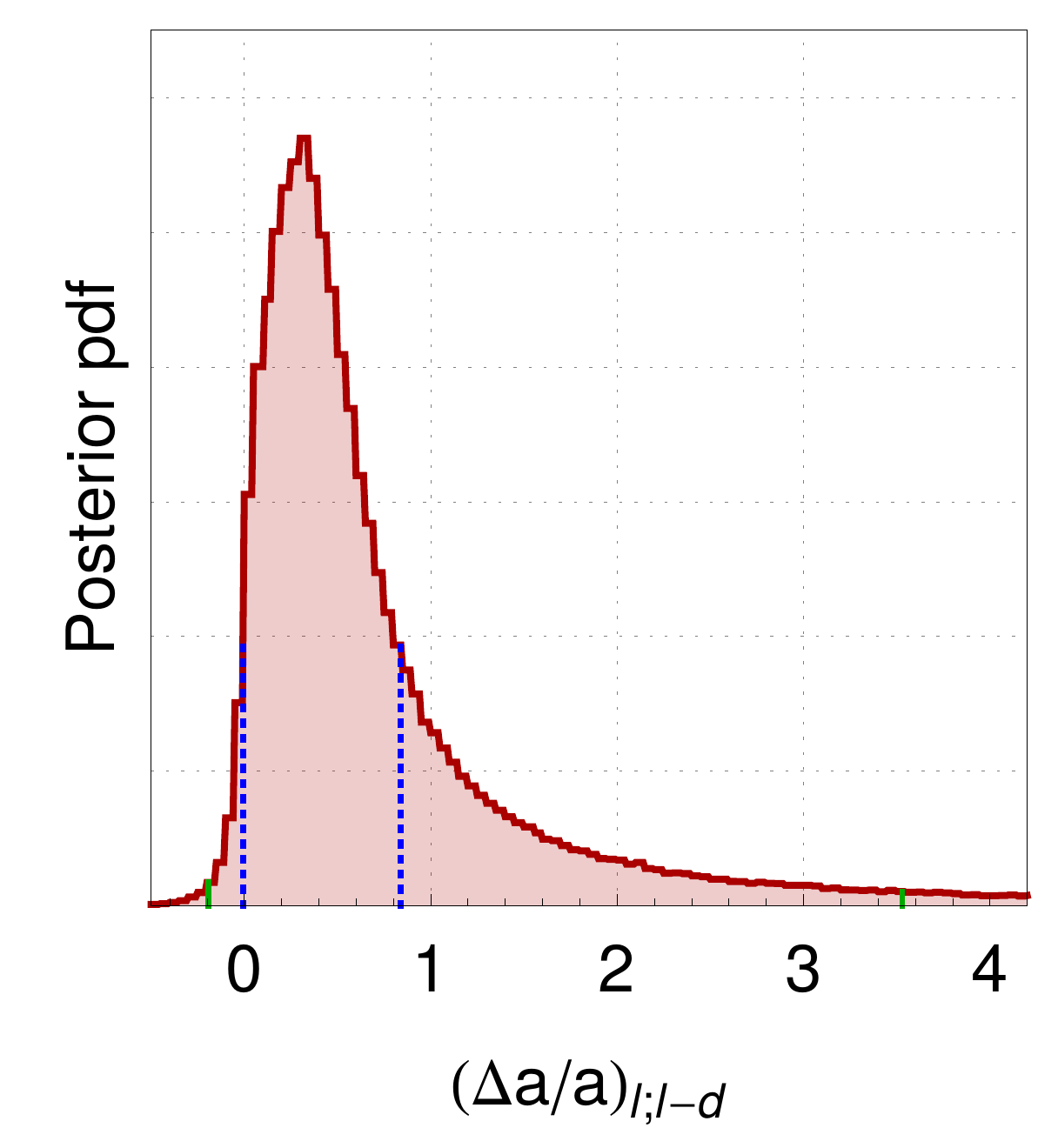}\hfill
\caption{Representation of the 1D marginalized posterior pdf inferred by our Bayesian analysis on the combinations $\alpha^2_{l,\infty}$, $\alpha_{d,\infty}\alpha_{l,\infty}$ as well as the relative differential acceleration given by Eq.~(\ref{eq:deltaa_a}) obtained using only galactic observations. The vertical dashed blue lines represent the 68 \% Bayesian credible intervals while the vertical continuous green lines represent the 95 \% Bayesian credible intervals.  We emphasize here that the posterior pdf on $(\Delta a/a)_{l;l-d}$ is computed using Eq.~(\ref{eq:deltaa_a}) and corresponds therefore to an indirect analysis that is model dependent.}
\label{fig:hist_pi}
\end{figure}

Instead of using the coupling strengths as the fundamental variables in our analysis, one can use the two variables $\alpha_{l,\infty}^2$ and $\alpha_{l,\infty}\alpha_{d,\infty}$. Using these two combinations is motivated by the fact that the observables depend explicitly on these two combinations and by the fact that the credible regions in the $\left(\alpha_{l,\infty},\alpha_{d,\infty}\right)$ plane have a ``hyperbolic-like'' shape (characterized approximately  by $\alpha_{l,\infty}\alpha_{d,\infty}=$ cst). An independent MCMC run using $\alpha_{l,\infty}^2$ and $\alpha_{l,\infty}\alpha_{d,\infty}$ leads to the credible regions shown on the right of Fig.~\ref{fig:conf_2D}. The shape of the credible regions is triangular and characterized by negative values of $\alpha_{l,\infty}\alpha_{d,\infty}$, which can also be seen on the histograms from Fig.~\ref{fig:hist_pi}. The main advantage of using these variables in the Bayesian inference comes from the fact that the credible region does not extend to very large values. The 1D histograms on these two parameters are presented on Fig.~\ref{fig:hist_pi}. The 68\% and 95\% (marginalized) credible intervals are presented in Tab.~\ref{tab:conf_pi}.

\begin{table}[htb]
\caption{Estimation of the 68\% and 95\% marginal credible intervals obtained using galactic observations alone.}\vspace*{2mm}
\label{tab:conf_pi} 
\centering
\begin{tabular}{c c c }
\hline
Parameter & 68\%  & 95\%  \\
\hline
$\alpha_{l,\infty}^2$ & $[0\, , \,0.023]$ & $[0 \, , 0.063]$ \\
$\alpha_{l,\infty}\alpha_{d,\infty}$ &[-0.67\, , 0.0]&$[-1.91\, ,\, 0.17]$ \\
$\left(\Delta a/a\right)_{l;l-d}$&$[-0.004\, ,\,  0.84]$ & $[-0.19\, ,\, 3.53]$ \\
\hline
\end{tabular}
\end{table}

As mentioned in the introduction, the model considered here can produce a violation of the EPP when the coupling strengths are not equal. This will result in a violation of the UFF. As we have seen in the previous paragraph, our analysis favours a region of the parameter space characterized by coupling strengths of opposite signs, thus leading to a violation of the UFF. It is interesting to quantify this violation. Using Eq.~(\ref{eq:deltaa_a}), we transform the results of our MCMC samplings into a sampling of $(\Delta a/a)_{l;l-d}$, the relative differential acceleration between a  test body made of DM and a test body made of SM falling in the same gravitational field generated by a body made of SM. The marginal 1D posterior pdf on $(\Delta a/a)_{l;l-d}$ is shown on the right of Fig.~\ref{fig:hist_pi}. No violation of the UFF is favoured by galactic observations at the 68\% credible level. The values of the credible intervals are also mentioned in Tab.~\ref{tab:conf_pi}. We emphasize here that the analysis on $(\Delta a/a)_{l;l-d}$ is an indirect analysis that is model dependent and is not a direct observation of $\Delta a/a$.

\subsubsection{Combined analysis using Solar System and galactic observations}

\begin{figure}[htb]
\includegraphics[width=0.45\textwidth]{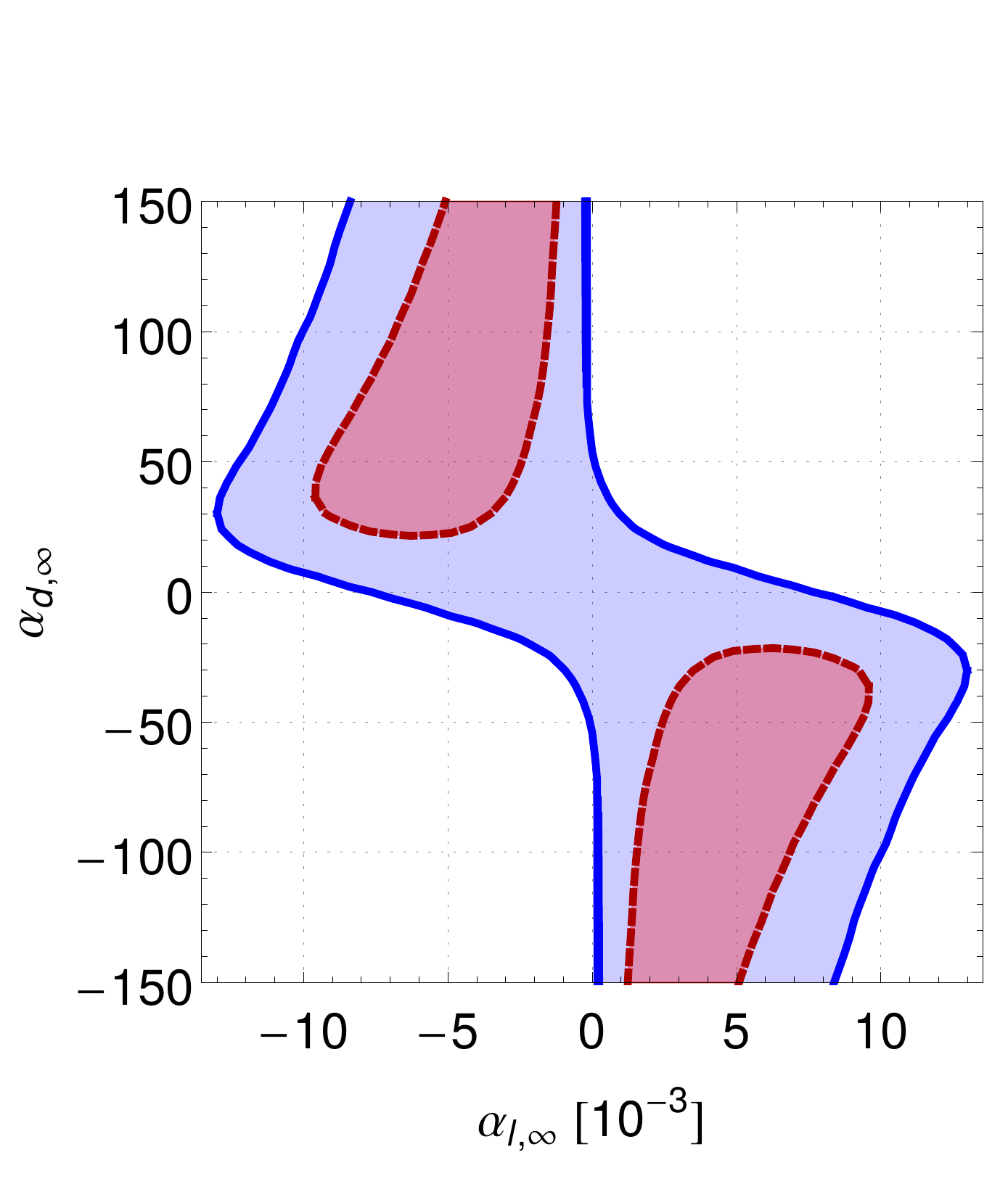}\hfill
\includegraphics[width=0.45\textwidth]{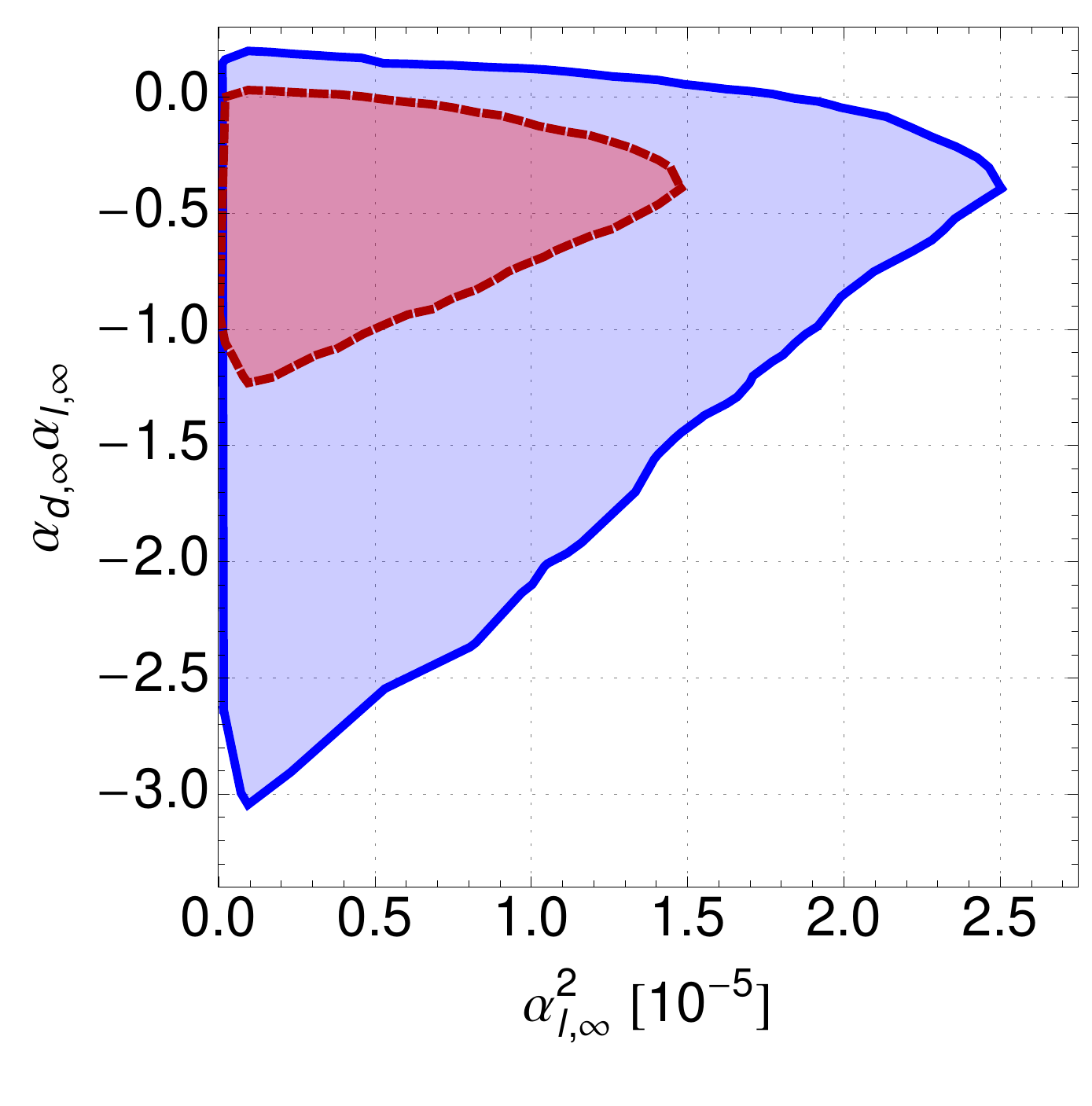}
\caption{Representation of the 68 \% (red, dashed lines) and 95 \% (blue, continuous line) credible region obtained using galactic observations combined with the Cassini measurement from~\citep{bertotti:2003uq}.}
\label{fig:conf_2D_Cas}
\end{figure}

\begin{figure}[htb]
\includegraphics[width=0.32\textwidth]{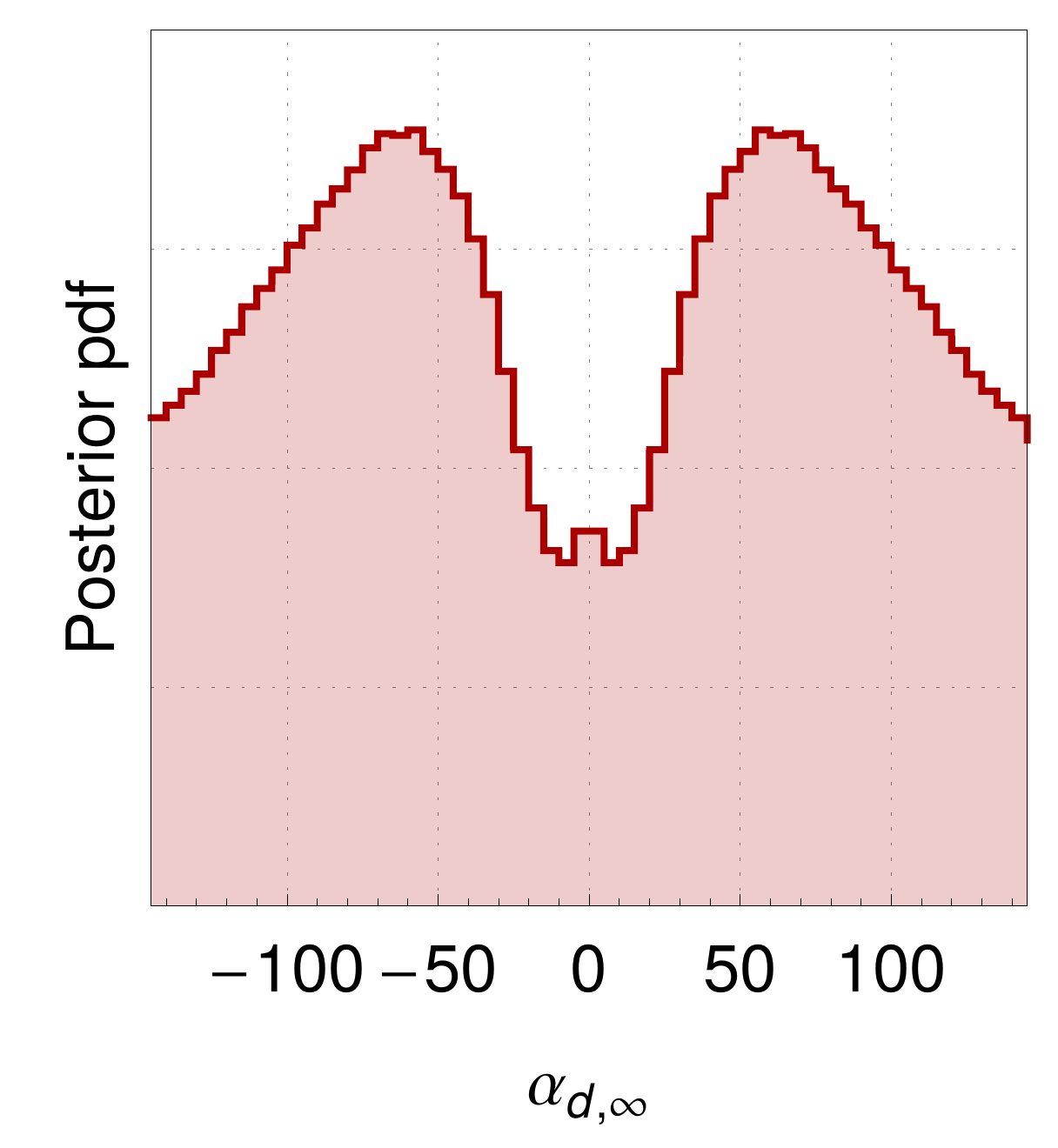}\hfill
\includegraphics[width=0.32\textwidth]{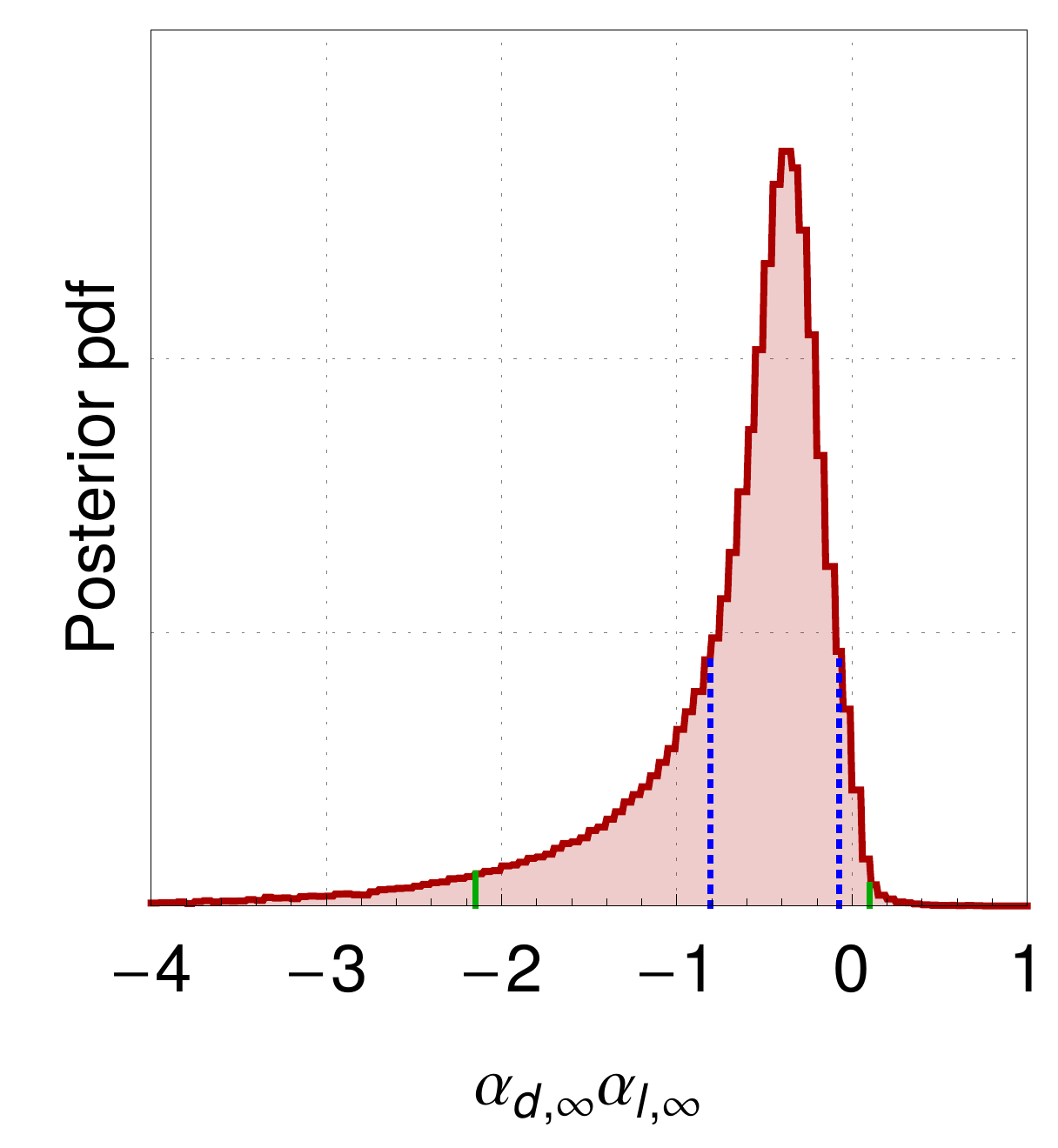}\hfill
\includegraphics[width=0.32\textwidth]{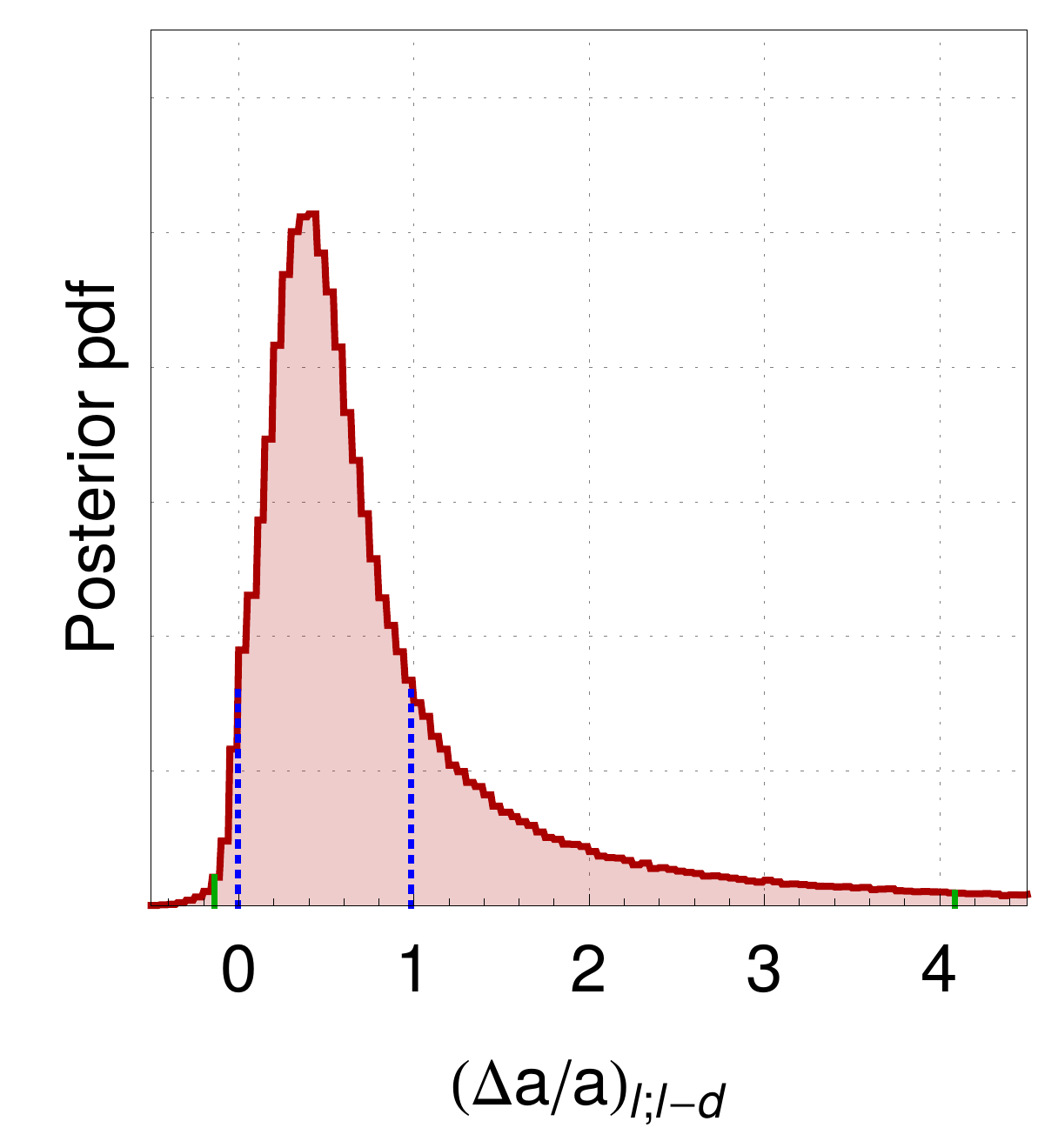}\hfill
\caption{Representation of the 1D marginalized posterior pdf inferred by our Basesian analysis on the combinations $\alpha^2_{l,\infty}$, $\alpha_{d,\infty}\alpha_{l,\infty}$ as well as the relative differential acceleration given by Eq.~(\ref{eq:deltaa_a}) obtained using galactic observations combined with the Cassini measurement from~\citep{bertotti:2003uq}. The vertical dashed blue lines represent the 68 \% credible interval while the vertical continuous green lines represent the 95 \% credible interval. We emphasize here that the posterior pdf on $(\Delta a/a)_{l;l-d}$ is computed using Eq.~(\ref{eq:deltaa_a}) and corresponds therefore to an indirect analysis that is model dependent.}
\label{fig:hist_Cas}
\end{figure}

As explained in Sec.~\ref{sec:priors}, the parameter $\alpha_{l,\infty}$ can be linked to the PPN parameter $\gppn$ in the Solar System (assuming DM to be negligible in the Solar System). Here, we will include the Solar System constraint on $\gppn$ coming from the Cassini spacecraft measurement~\cite{bertotti:2003uq}. More precisely, we produce a new MCMC run using a prior pdf on $\alpha_{l,\infty}$ derived from the Cassini constraint (see the discussion in Sec.~\ref{sec:priors}). Fig.~\ref{fig:conf_2D_Cas} represents the 2D marginal credible regions obtained including the Cassini constraint. These two figures can be compared to the ones from Fig.~\ref{fig:conf_2D}. One can see that the regions are much smaller in the $\alpha_{l,\infty}^2$ direction, which shows the strong impact of the Solar System constraint. Nevertheless, no improvement can be noticed on the $\alpha_{d,\infty}\alpha_{l,\infty}$ direction. This is confirmed by comparing the 1D histogram from the middle of Fig.~\ref{fig:hist_Cas} with the middle of Fig.~\ref{fig:hist_pi}: both distributions are very similar (see also the values of the credible intervals mentioned in Tab.~\ref{tab:conf_pi} and in Tab.~\ref{tab:conf_pi_cass}). As a consequence, the indirect constraint on $(\Delta a/a)_{l;l-d}$ is not improved by the addition of the Cassini constraint. This can be seen from the right histograms of Fig.~\ref{fig:hist_Cas} and~\ref{fig:hist_pi} (see also the values of the credible intervals mentioned in Tab.~\ref{tab:conf_pi} and in Tab.~\ref{tab:conf_pi_cass}).

The fact that $\alpha_{l,\infty}\alpha_{d,\infty}$ remains of the same order of magnitude as previously while the quantity $\alpha_{l,\infty}^2$ is reduced by a factor $10^4$ implies  that estimations of $\alpha_{d,\infty}$ will be much larger than previously. This can be seen from the credible regions on the left panel of Fig.~\ref{fig:conf_2D_Cas}. This is also confirmed by the 1D marginal posterior pdf on the parameter $\alpha_{d,\infty}$ that is shown on the left of Fig.~\ref{fig:hist_Cas}. This is an interesting fact: the Cassini constraint on the $\gppn$ parameter implies that estimations of $\alpha_{l,\infty}$ are much smaller but also lead to higher estimations in the parameter $\alpha_{d,\infty}$. Let us recall that our modelling is valid only up to values of $\alpha_{d,\infty}$ of the order of $10^3$.

\begin{table}[htb]
\caption{Estimation of the 68\% and 95\% marginal credible intervals obtained using galactic observations combined with Solar System PPN constraint.}\vspace*{2mm}
\label{tab:conf_pi_cass} 
\centering
\begin{tabular}{c c c }
\hline
Parameter & 68\%  & 95\%  \\
\hline
$\alpha_{l,\infty}\alpha_{d,\infty}$ &$[-0.81\, ,\, -0.07]$&  $[-2.15\, ,\, 0.10]$ \\
$\left(\Delta a/a\right)_{l;l-d}$ &$[-0.003\, ,\, 0.98]$&  $[-0.14\, ,\, 4.09 ]$ \\
\hline
\end{tabular}
\end{table}

\subsection{Universal coupling}\label{sec:BD}
\begin{figure}[htb]
\centering
\includegraphics[scale=0.4]{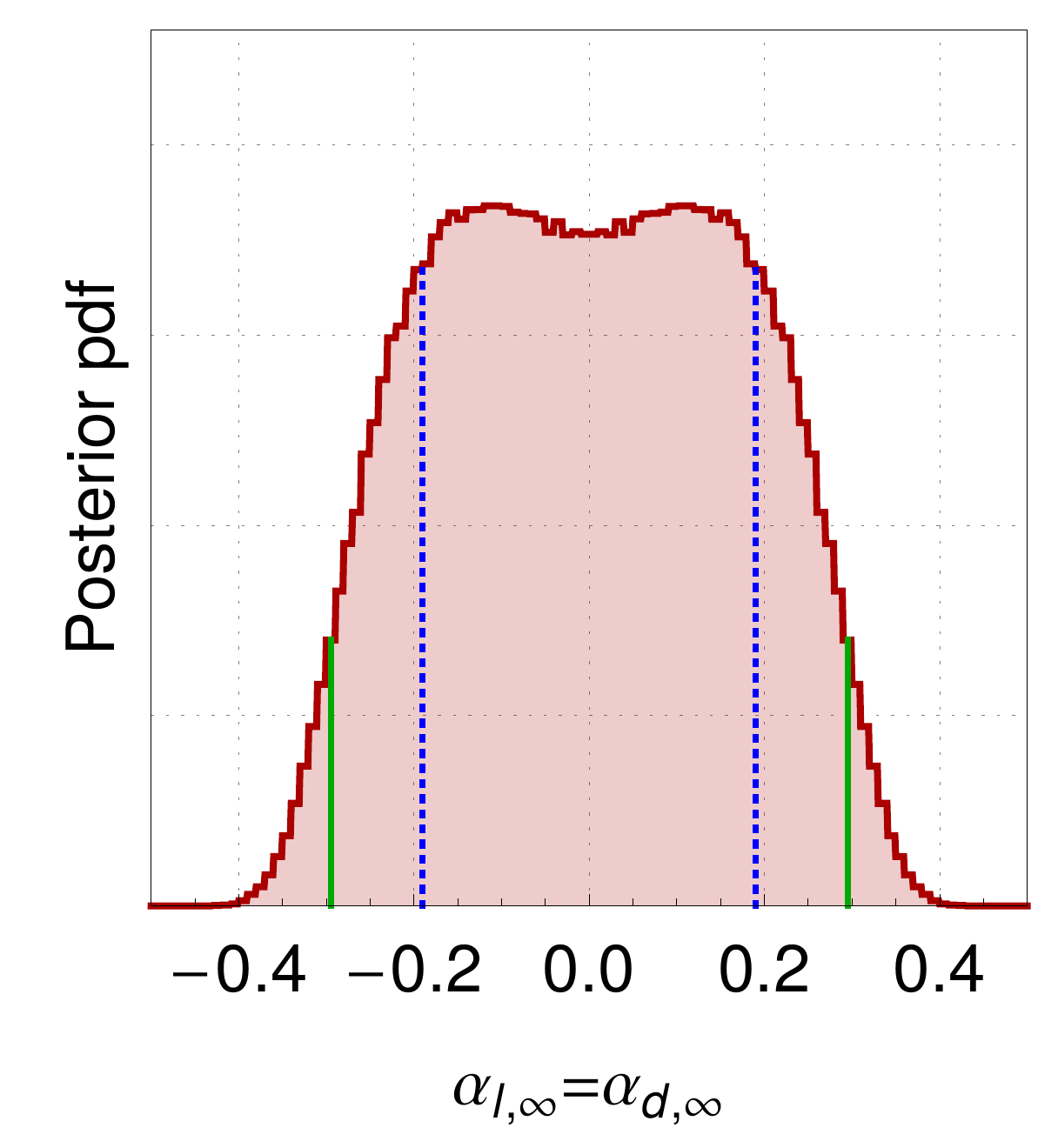}\hspace{0.12\linewidth}
\includegraphics[scale=0.4]{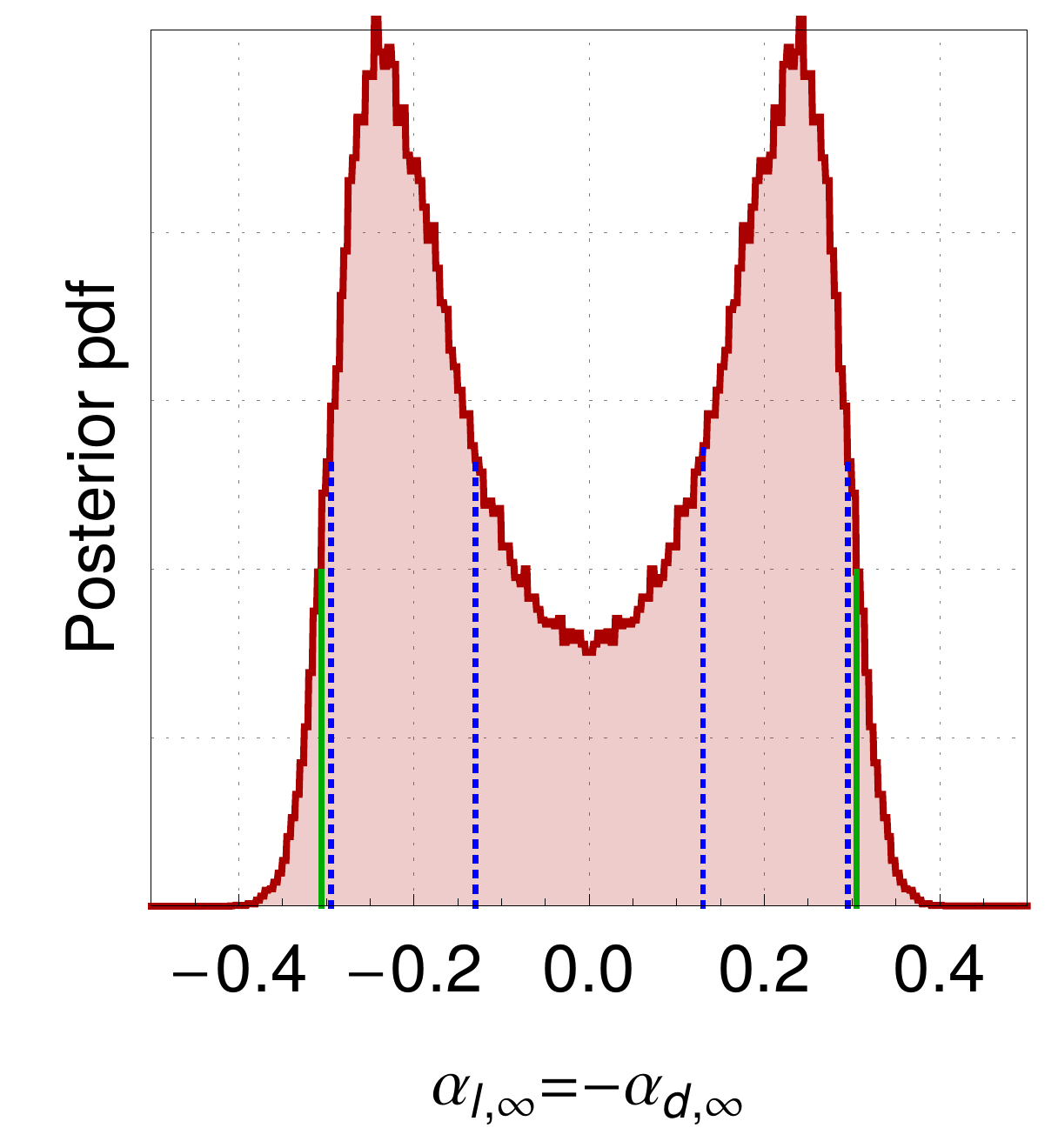}
\caption{Representation of the 1D marginalized posterior pdf inferred by our Bayesian analysis on the coupling strength $\alpha_{l,\infty}$ obtained using only galactic observations. Left: Brans-Dicke theory characterized by of a universal coupling ($\alpha_{l,\infty}=\alpha_{d,\infty}$). Right: opposite coupling strengths ($\alpha_{l,\infty}=-\alpha_{d,\infty}$). The vertical dashed blue lines represent the 68 \% credible interval while the vertical continuous green lines represent the 95 \% credible interval.}
\label{fig:hist_BD_AWE}
\end{figure}

In this section, we will enforce the coupling strengths to be equal $\alpha_{l,\infty}=\alpha_{d,\infty}$. In this case, we have a universal coupling between the scalar field and matter (in the Einstein frame) which corresponds to a case with no violation of the EEP. This can be seen from the action~(\ref{action JF}) where $M(\Phi)=1$ in the case of equal coupling functions. We are therefore considering a standard Brans-Dicke theory~\citep{jordan:1949vn,brans:1961fk,brans:2014sc}. The marginal posterior pdf on the coupling strength $\alpha_{l,\infty}$ is shown on the left of Fig.~\ref{fig:hist_BD_AWE}. It has a plateau between -0.2 and 0.2 and rapidly goes to zero on either side of this plateau. The related Bayesian credible intervals are presented in Tab.~\ref{tab:conf_BD}. This constraint can be transformed into a constraint on the $\gppn$ parameter using the Eq.~(\ref{eq:gppn}). This leads to a negative estimation of $\gppn-1$ (logical since this kind of tensor scalar theories can only produce negative values of $\gppn-1$). The 68\% credible interval on the $\gppn$ parameter is given by $[-0.07,0]$. A constraint at the level of $10^{-2}$ on this parameter has also been obtained in~\cite{bolton:2006uq,schwab:2010fk} by using a very similar dataset but another method.

In the case of a pure Brans-Dicke theory (characterized by the action~(\ref{action JF}) with $h(\Phi)=\Phi$, $M(\Phi)=1$ and $\omega(\Phi)=\omega_\textrm{BD}$), the $\alpha_{l,\infty}$ parameter is related to the $\omega_\textrm{BD}$ parameter through $2\alpha_{l,\infty}^2/(1+\alpha_{l,\infty}^2)=(2\omega_\textrm{BD}+3)^{-1}$. The 1-$\sigma$ credible interval on $\alpha_{l,\infty}^2$ leads to an estimation of $\omega_\textrm{BD}>28.9$. The best constraint is  $\omega_{BD} > 4 \times 10^{4}$ coming from the Cassini measurement~\cite{bertotti:2003uq}. Note that binary systems also provided constraints at the level of $\omega_\textrm{BD} >  1250$~\cite{will:1989qy,alsing:2012qy}.

\begin{table}[htb]
\caption{Estimation of the 68\% and 95\% credible intervals of the coupling strength obtained using galactic observations.}\vspace*{2mm}
\label{tab:conf_BD} 
\centering
\begin{tabular}{ c c c c }
\hline
& Parameter & 68\%  & 95\%  \\
\hline
Brans-Dicke (universal coupling) & $\alpha_{l,\infty}=\alpha_{d,\infty}$ & $[-0.19 \, , 0.19]$ & $[-0.295 \, , 0.295]$ \\
Opposite couplings & $\alpha_{l,\infty}=-\alpha_{d,\infty}$ & $\pm [0.128 \, , 0.295]$ & $[-0.304 \, , 0.304]$ \\
\hline
\end{tabular}
\end{table}

\subsection{Coupling strengths of same magnitude but opposite signs}\label{sec:AWE}

In this section, we will enforce the coupling strengths to be of the same magnitude but with opposite signs $\alpha_{l,\infty}=-\alpha_{d,\infty}$. This case is known to have interesting properties at cosmological scales: it allows one to mimic Dark Energy and produces an interesting explanation to the cosmological coincidence problem~\citep{fuzfa:2007mw,alimi:2008zr}.  This case leads to a violation of the UFF characterized by $(\Delta a/a)_{l;l-d}=2\alpha_{l,\infty}^2$. The marginal posterior pdf of $\alpha_{l,\infty}$ is shown on the right of Fig.~\ref{fig:hist_BD_AWE} and the related credible intervals are in Tab.~\ref{tab:conf_BD}. The data also seems to indicate a violation of the EEP at 1$\sigma$ level (but everything is compatible with the EEP at 2$\sigma$).

\subsection{Influence of the Hubble and deceleration cosmic parameters on the results}\label{sec:influence_cosmo}
\begin{figure}[htb]
	\includegraphics[width=0.53\textwidth]{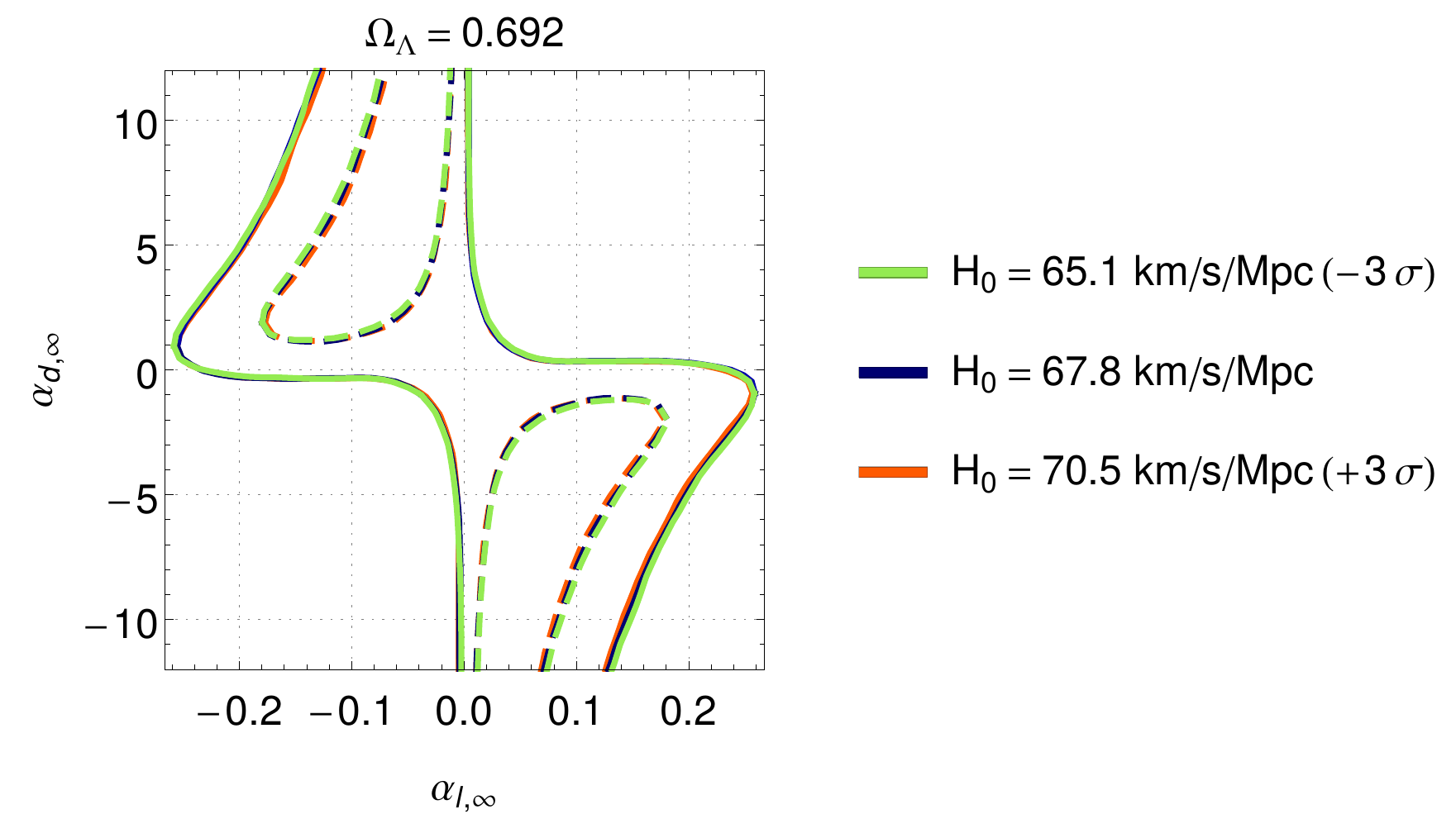}\hfill
	\includegraphics[width=0.46\textwidth]{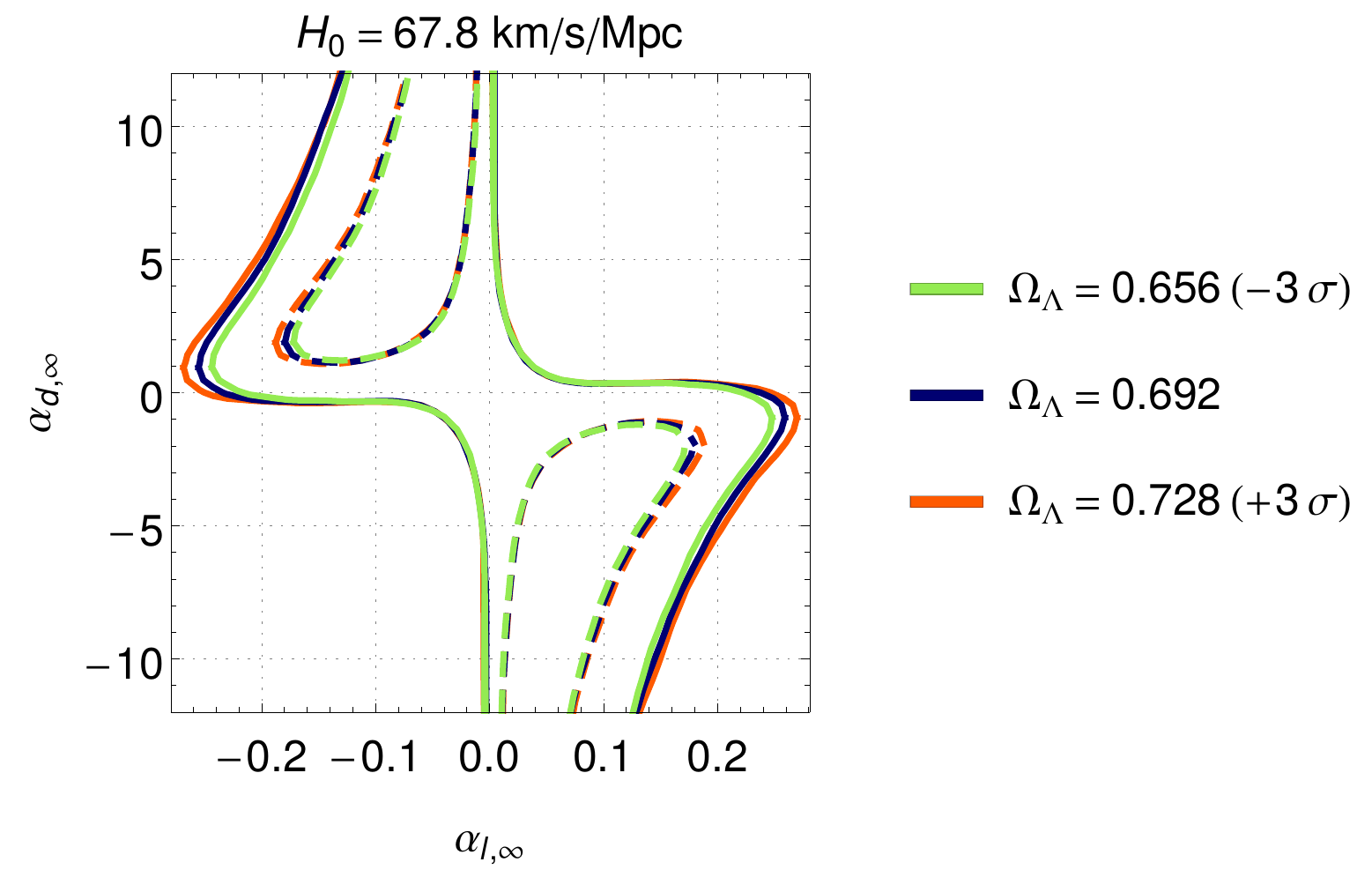}
	\caption{The effect of varying $H_0$ and $\Omega_\Lambda$ on the 68 \% (dashed lines) and the 95 \% (continuous lines) on the credible regions of independent $\alpha_{l, \infty}$ and $\alpha_{d, \infty}$. Only galactic observations are used for the above constraints. The leftmost plot is for a fixes $\Omega_\Lambda$ while the rightmost plot fixes $H_0.$ In both cases we consider a variation of $\pm 3 \sigma$ for the variable cosmological parameter.}
\label{fig:conf_2D_imposed}
\end{figure}

The deflection angle is dependent on the Hubble rate and the cosmological constant through angular diameter distances. As a consequence our analysis may be sensitive to the values we have chosen for $H_0$ and $\Omega_\Lambda$. In this paper, the cosmological values $H_0 = 67.8 \mbox{km/Mpc/s}$ and $\Omega_\Lambda = 0.692$ obtained by Planck were chosen~\cite{2015arXiv150201589P}. Studies that use lensing to estimate cosmological parameters \cite{bolton:2006uq,2008A&A...477..397G,schwab:2010fk,2012ApJ...755...31C} find values of $\Omega_\Lambda$ consistent with our chosen fiducial values. The margin of error on the cosmological parameters using galaxy-galaxy lensing remains large. Using the best available constraints $H_0 =  67.8 \pm 0.9 \, \mbox{km/Mpc/s} \, , \, \Omega_\Lambda = 0.692 \pm 0.012$ \cite{2015arXiv150201589P}, one can show that the effect of the cosmological parameters on the estimated credible region for the coupling strengths is weak. This is illustrated in Fig. \ref{fig:conf_2D_imposed} in which credible regions on the estimated parameters obtained using different values for the cosmological parameters are shown. One can see that the credible region does not change significantly when changing the cosmological parameters.  Similar results hold in case of a single independent coupling and in the combined Solar System and galactic observations cases. A more robust analysis would encode what is known about $H_0$ and $\Omega_\Lambda$ as priors and marginalize over these parameters to obtain results independent of these parameters. However, the analysis presented in this section indicates there is little advantage to be gained by this and that the results would not change significantly. 

\section{Conclusion}\label{sec:conclusion}

In this paper, we presented a general framework to test the Einstein Equivalence Principle (EEP) in the Dark sector on galactic scales. To this end, we use combined observational constraints on strong lensing and velocity dispersions within the lens galaxies. The dataset used is a sample of 53 galaxies for which we have both lensing and velocity dispersion data available. We modelled potential violations of the EEP via a simple scalar-tensor theory non-universally coupled to standard luminous matter (SM) and Dark Matter (DM). The departure from GR is parametrized by 2 coupling strengths $\alpha_{l,\infty}$ and $\alpha_{d,\infty}$ (derivatives of the logarithm of the couplings  to SM $\alpha_{l,\infty}=\frac{d \ln A_l \left( \varphi \right)}{d \varphi}|_{\varphi_{\infty}}$ and to DM $\alpha_{d,\infty}=\frac{d \ln A_d \left( \varphi \right)}{d \varphi}|_{\varphi_{\infty}}$). By employing a weak field approximation, we derived analytical expressions, valid at leading order in the value of the scalar field, for both the radius of the Einstein ring of lens galaxies and the velocity dispersion in these lens galaxies. We showed that the relevant parameters actually constrained by these observations were the combinations of the coupling strengths $\alpha_{l,\infty}^{2}$ and $\alpha_{l,\infty}\alpha_{d,\infty}$ whose constraints are presented on Fig.~\ref{fig:conf_2D}. As a result of this degeneracy, while constraints on the coupling strength to SM are reasonably strong (see left of Fig.~\ref{fig:hist_alpha}), the coupling strength to DM is very poorly constrained (see right of Fig.~\ref{fig:hist_alpha}). Combining galactic observations with Solar System constraints significantly improves the constraints on the coupling strength to SM, but the coupling strength to DM remains poorly (if at all) constrained (see left of Fig.~\ref{fig:hist_Cas}); actually, because of the degeneracy mentioned above, constraints on $\alpha_{d,\infty}$ are relaxed by the introduction of the Solar System constraints.

Probably the most interesting result of this analysis is that that data favour coupling strengths of opposite signs (see middle of Fig.~\ref{fig:hist_pi}): if any violation of the EEP is real, it is more likely to present itself with coupling strengths of opposite signs. Whether or not this preference has a dynamical origin is still unclear and will be investigated in a future work, but it might be an interesting hint about non-universally coupled theories of gravity.
The coupling strengths can be used as an indirect probe of the Universality of Free Fall (UFF). We find that the inferred anomalous acceleration between SM and DM falling in the gravitational field of SM $\left(\frac{\Delta a}{a}\right)_{l;l-d}$ is compatible with GR at 1-$\sigma$ and has a very long non-Gaussian tail, thus preventing us from ruling out potential violation of the UFF. We emphasize here again that constraints on the UFF are indirect since coming from Eq.~(\ref{eq:deltaa_a}) and are therefore model dependent. 
 
 These results indicate that, in order to lift the degeneracy and to be able to probe $\alpha_{d,\infty}$ efficiently, one should turn to observables dominated by the dynamics of DM. Therefore, the next step in testing the UFF between ordinary and DM will be to study the effects of the anomalous couplings on cosmological structure formation in the linear regime. In particular, in these models, we expect both a change in the linear growth rate of large scale structure and an additional bias between the power spectrum of DM and the observed power spectrum of galaxies on large scales. Such quantities could be constrained in the near future with large scale surveys such as the SKA, using, e.g. redshift space distortion \cite{Maartens:2015mra,Zhao:2015wqa,Raccanelli:2015qqa}.

\acknowledgments The authors thank A. F\"uzfa and B. Famaey for interesting discussions on this topic, C. Stubbs for pointing interesting references to them and the anonymous referee for useful comments that helped to improve the quality of the publication. AH thanks A. Rivoldini for useful discussions about the Bayesian inference. JL and NM are supported by the National Research Foundation (South Africa).

\newpage
\appendix
\section{Field equations and weak field solutions}\label{app:WF}
\subsection{Einstein frame and field equations}\label{app:einstein}

In Sec.~\ref{Intro}, we discuss what form the gravitational action may take in theories in which the EEP may not hold. In particular we consider scalar-tensor theories given by the Jordan Frame action (see Eq.~\ref{action JF} )

\begin{align}
	S = \frac{c^4}{16\pi G_*}\int d^4x \sqrt{-g}\left[h(\Phi)R-\frac{\omega(\Phi)}{\Phi}g^{\mu\nu}\partial_\mu\Phi\partial_\nu\Phi\right] + S_{l}\left[g_{\mu\nu},\Psi_{l}\right]+S_{d}\left[M^2(\Phi)g_{\mu\nu},\Psi_{d}\right],
\end{align}
where the subscripts $l$ and $d$ refer to SM and DM respectively. We can move from the Jordan frame to the Einstein frame by applying the following conformal transformations and scalar field redefinition:

\begin{subequations}\label{eq:conf_transf}
\begin{align}
g_{\mu \nu} &= A^2_l(\varphi)g^*_{\mu \nu} \, ,\\
A_{l}^2(\varphi) &= h^{-1}\left(\Phi \right)\, , \\
\left( \frac{d \varphi}{d \Phi}\right)^2 &\equiv \frac{3}{4} \left( \frac{d \ln h\left( \Phi \right)}{ d \Phi}\right)^2 + \frac{\omega \left( \Phi \right)}{2 \Phi h \left( \Phi \right)} \, ,\\
 M(\Phi) &\equiv \frac{A_{d} \left( \varphi \right) }{A_{l} \left( \varphi \right)} \, .
\end{align}
\end{subequations}
The action then takes the form
\begin{align}
	S&=\frac{c^4}{16 \pi G_*}\int d^4x \sqrt{-g_*}\left[R_*-2 g^{\mu\nu}_*\partial_\mu\varphi\partial_\nu\varphi\right] +S_{l}\left[A^2_{l}(\varphi)g^*_{\mu\nu},\Psi_{l}\right]+S_{d}\left[A^2_{d}(\varphi)g^*_{\mu\nu},\Psi_{d}\right] \, ,
\end{align}
where the stars indicate quantities expressed in the Einstein frame. In both frames the bare gravitational constant, $G_*$, can be related to the observed gravitational constant $G$ by the formula:
\begin{align}\label{eq:G_obs}
G & = G_* A_l^2 \left( \varphi\right)  = \frac{G_*} {h\left( \Phi \right)}\, .
\end{align}

From the variation of the action we are able to derive the following field equations
\begin{subequations}\label{eq:field_einstein}
\begin{align}
R_{\mu \nu}^* - \frac{1}{2}R^* g_{\mu \nu}^* &= \frac{8 \pi G_*}{c^4} \left( T^{*({l})}_{\mu \nu} + T^{*({d)}}_{\mu \nu} \right) +  2 \partial_{\mu} \varphi \partial_{\nu} \varphi - g_{\mu \nu}^* \left( g^{\alpha \beta}_* \partial_\alpha \varphi \partial_\beta \varphi \right) \, ,\\
\Box_* \varphi &= -\frac{4 \pi G_*}{c^4} \left( \alpha_{l}(\varphi) T^{(l)}_* + \alpha_{d}(\varphi) T_*^{(d)}\right) \, ,
\end{align}
\end{subequations}
where the two energy momentum tensors are given by
\begin{equation}
	T^{*(l)}_{\mu \nu} =-  \frac{2}{\sqrt{-g_*}} \frac{\delta S_{l}}{\delta g_*^{\mu \nu}} \, , \qquad 	T^{*(d)}_{\mu \nu} =-  \frac{2}{\sqrt{-g_*}} \frac{\delta S_{d}}{\delta g_*^{\mu \nu}} \, .
\end{equation}
Similarly, the invariance under diffeomorphism lead to the conservation equations
\begin{equation}
	\nabla^*_\mu T^{(l) \ \mu}_\nu=  \alpha_l(\varphi) T^{(l)}_* \nabla^*_\nu \varphi \, , \qquad 	\nabla^*_\mu T^{(d) \ \mu}_\nu=  \alpha_d(\varphi) T^{(l)}_* \nabla^*_\nu \varphi \, .
\end{equation}
In the last equation, we suppose that there is no interaction between DM and SM except through the scalar field (see also Eq.~(9) from \cite{damour:1990fk} or Eq.~(6) from \cite{alimi:2008zr}).

In the preceding equations, the running of the coupling $\alpha_{l}$ and $\alpha_d$ are defined by
\begin{align}
\alpha_{l} \left( \varphi \right) = \frac{d \ln A_{l} \left(\varphi \right)}{d \varphi}\, , \qquad  \alpha_{d} \left( \varphi \right) = \frac{d \ln A_{d} \left(\varphi \right)}{d \varphi} \, .
\end{align}
We assume both types of matter to be described by a perfect fluid. The expression of the Einstein frame stress energy tensor is given by 
\begin{equation}
	T^{*(i)}_{\mu \nu} = \left(\rho^*_{(i)}c^2 + p^*_{(i)} \right)u^*_{(i)\mu} u^*_{(i)\nu} + p^*_{(i)}g^*_{\mu \nu}\, ,
\end{equation}
where $\rho^*_{(i)}$ and $p^*_{(i)}$ are the matter density and pressure and $u^*_{(i)\mu}$ is the 4-velocity of the fluid,  of the type of matter $i$ (SM or DM). These Einstein frame stress energy tensors are related to their Jordan frame counterparts by the relation~\cite{damour:1992ys}
\begin{equation}\label{eq:conservation}
		T^{*(i)\mu}_{ \nu}=A^4_i(\varphi)T^{(i)\mu}_{ \nu}\, ,
\end{equation}
with $T^{(i)\mu}_{ \nu}= \left(\rho_{(i)}c^2 + p_{(i)} \right)u_{(i)\mu} u_{(i)\nu} + p_{(i)}g_{\mu \nu}$ where $\rho_{(i)}$ and $p_{(i)}$ are the observable matter density and pressure. This implies the relation between matter density and pressure between the two frames~\citep{damour:1992ys,damour:1993kx,damour:1993uq}
\begin{subequations}
	\begin{align}
		\rho^*_{(i)}&=A^4_l(\varphi)		\rho_{(i)}\, ,\\
		p^*_{(i)}&=A^4_l(\varphi)	p_{(i)}\,.
	\end{align}
\end{subequations}

\subsection{Spherical symmetry}

Assuming spherical symmetry and a static configuration, one can write the Einstein frame metric in isotropic coordinates as
\begin{equation}
	ds^2_*=-e^{\nu_*}c^2dt^2_* + e^{\lambda_*}\left(dr_*^2+r_*^2d\Omega^2\right)\, ,
\end{equation}
where $\nu_*$ and $\lambda_*$ depends only on the radial coordinate $r_*$. Introducing this expression in the field equations~(\ref{eq:field_einstein}) and in the conservation equation~(\ref{eq:conservation}) leads to 

\begin{subequations}\label{eq:spherical_field}
	\begin{align}
		-\lambda_*''-\frac{2}{r_*}\lambda_*'&= \frac{8\pi G_*}{c^2} e^{\lambda_*}A^4_l(\varphi)\rho_t + \varphi'^2 +\frac{\lambda_*'^2}{4}\, \label{lam_FE},\\
	\lambda_*' + \nu_*' &=  \frac{8\pi G_*}{c^4} e^{\lambda_*} A^4_l(\varphi) r_* p_t + r_* \varphi'^2 -r_*\frac{\nu_*' \lambda_*'}{2}- r_*\frac{\lambda_*'^2}{4} \, , \\	
	\nu''_* +\frac{2\nu'_*}{r_*} &= \frac{8 \pi G_*}{c^4} e^{\lambda_*} A^4_l(\varphi)\left( \rho_t c^2 +p_t\right)-\frac{\nu_*'^2}{2} - \frac{\lambda_*' \nu_*'}{2}\, , \\
	\varphi'' + \varphi' \left( \frac{\lambda_*' + \nu_*'}{2} + \frac{2}{r_*}\right)&= \frac{4 \pi G_*}{c^4} e^{\lambda_*} A^4_l(\varphi)\left[ \alpha_l(\varphi)  \left( \rho_l c^2 - 3 p_l \right) + \alpha_d(\varphi) \left( \rho_d c^2 - 3 p_d\right)\right] \, , \label{eq:kg_eq}\\
	p'_{l} &= -  \left(\alpha_l(\varphi)\varphi'+\frac{\nu_*'}{2}\right)  \left( \rho_{l}c^2 +3p_{l}  \right)\, ,\\
	p'_{d} &= -  \left(\alpha_d(\varphi)\varphi'+\frac{\nu_*'}{2}\right)  \left( \rho_{d}c^2 +3p_{d}  \right)\, ,
	\end{align}
\end{subequations}
where 
\begin{equation}
	\rho_t=\rho_l+\rho_d \, \quad \textrm{and} \quad p_t=p_l+p_d \, .
\end{equation}
The prime represents the derivative with respect to $r_*$. These equations need to be complemented by a set of boundary conditions and an equation of state relating the pressures and the densities. The equations can be solved numerically similarly to what is done in~\cite{damour:1993vn} (note that in that paper, they work in Schwarzschild coordinates whereas we are working in isotropic coordinates which are more suitable to the weak field observables). In the following, we will use a weak field approximation of the previous system of equations.

\subsection{Weak field limit in the Einstein frame}\label{app:weak_einstein}

The galactic gravitational field is always very small. Therefore, it is justified to use a weak field expansion of the Eqs.~(\ref{eq:spherical_field}) which allows one to write the metric components as 
\begin{subequations}
\begin{align}
e^{\nu_*} &\approx 1 + \nu_* \, ,\\ 
e^{\lambda_*} &\approx 1 + \lambda_*\, . 
\end{align}
\end{subequations}

Moreover, the pressure can always be neglected with respect to the density ($p/\rho c^2\ll 1$). Making this approximation, in the Einstein frame, results in a set of simpler equations that read
\begin{subequations}\label{eq:low_field1}
	\begin{align}
			\Delta_*\lambda_* &= \lambda_*'' +\frac{2}{r_*}\lambda_*' =-\frac{8\pi G_*}{c^2} A^4_{l}(\varphi)\rho_t  \, ,\\
			\Delta_*\nu* &= \nu_*'' +\frac{2}{r_*}\nu_*' =\frac{8\pi G_*}{c^2} A^4_{l}(\varphi)\rho_t  \, ,\\			
			\varphi''+\frac{2}{r_*}\varphi'&= -4\frac{\pi G_*}{c^2} A^4_{l}(\varphi)\left( \alpha_{l}(\varphi)\rho_l+\alpha_{d}(\varphi)\rho_d \right) \, .
	\end{align}
\end{subequations}
Note that in the previous equations, we have neglected terms of the form $\varphi'^2$. The condition under which we can neglect these terms can be determined a posteriori by using the solution of the scalar field given by Eq.~(\ref{eq:scal_sol}) and is given by $\alpha^2_{d} \Xi \ll 1$ where $\Xi$ is the compactness parameter $\Xi\sim G_*M_*/c^2r_*$. In a typical galaxy, $\Xi$ is of the order of $10^{-7}$ which means that our weak field approximation will break down for values of $\alpha_d$ of the order of $10^3$. Above this limit, a strong gravitational field will be created by the scalar field (itself generated by the strong coupling to the Dark sector). This strong field scenario is not considered here.

\subsection{Weak field limit in the Jordan frame for luminous matter}\label{app:weak_jordan}

The observables are more easily computed from the Jordan frame. The Jordan frame metric is related to the Einstein frame one through the conformal transformation~(\ref{eq:conf_transf}) which can be written in the weak scalar field limit as

\begin{equation}\label{eq:conf}
	ds^2=A^2_l(\varphi)ds^2_*= A^2_{l,\infty}\left[1+2\alpha_{l,\infty}(\varphi-\varphi_\infty)\right] ds^2_*\, ,
\end{equation}
where the subscripts $\infty$ refer to quantities evaluated at $\varphi_\infty=\varphi(r_*=\infty)$.
In order to have an asymptotically flat space-time metric, one should do a rescaling of the 4-coordinates \begin{equation}\label{eq:rescaling}
	x^\mu=A_{l,\infty}x^\mu_* \, .
\end{equation}
This rescaling implies in particular a rescaling of the radial coordinate:
\begin{equation}
	r=A_{l,\infty}r_*\,. 
\end{equation}
Using this coordinate in the weak field Eqs.~(\ref{eq:low_field1}) leads to
\begin{subequations}\label{eq:low_field2}
	\begin{align}
			\Delta\lambda_* &= \ddot\lambda_* +\frac{2}{r}\dot \lambda_* =-\frac{8\pi G}{c^2} \frac{A^4_l(\varphi)}{A^4_{l,\infty}}\rho_t(r) \, ,\\
			\Delta\nu_* &= \ddot \nu_* +\frac{2}{r}\dot \nu_* =\frac{8\pi G}{c^2}\frac{A^4_l(\varphi)}{A^4_{l,\infty}} \rho_t(r)  \, ,\label{eq:nu_low_field}\\			
			\ddot \varphi+\frac{2}{r}\dot \varphi&= -\frac{4\pi G}{c^2}\frac{A^4_l(\varphi)}{A^4_{l,\infty}} \left( \alpha_{l}(\varphi)\rho_l(r)+\alpha_{d}(\varphi)\rho_d(r) \right) \, ,\label{eq:scalar_low_field}
	\end{align}
\end{subequations}
where the dot represents the derivative with respect to $r$. Note that in the last expressions, we make use of the observed gravitational constant defined by Eq.~(\ref{eq:G_obs}). 

We introduce  a parametrization of the weak field Jordan frame metric in terms of standard gravitational potentials $\Psi$ and $\Phi$ of the form
\begin{equation}\label{eq:metric_WJF}
ds^2 =  -\left( 1 + 2\frac{\Psi (r)}{c^2}  \right)c^2 dt^2 + \left( 1 - 2 \frac{\Phi(r)}{c^2}\right) dl^2\, .
\end{equation}
Using the conformal transformation (\ref{eq:conf}) with the rescaling of the variables (\ref{eq:rescaling}), the gravitational potentials are given by
\begin{subequations}
	\begin{align}
		\frac{\Psi(r)}{c^2}&= \alpha_{l,\infty}(\varphi-\varphi_\infty)+\frac{\nu_*}{2}\, \label{Psi_eq},\\
		\frac{\Phi(r)}{c^2}&= -\alpha_{l,\infty}(\varphi-\varphi_\infty)-\frac{\lambda_*}{2}\, ,
	\end{align}
\end{subequations}
where $\varphi$, $\lambda_*$ and $\nu_*$ are solutions of the Eqs.~(\ref{eq:low_field2}). Note that, as we shall see below, the important quantity for the lensing is $\Psi+\Phi$ which is given by 
\begin{equation}
	\frac{\Psi+\Phi}{c^2}=\frac{\nu_*-\lambda_*}{2} \, ,
\end{equation}
so that
\begin{equation} \label{laplacian_pot}
	\Delta \left(\Psi+\Phi\right)=8\pi G\frac{A^4_l(\varphi)}{A^4_{l,\infty}}\rho_t(r) \, .
\end{equation}

\subsection{Weak field solution for the scalar field}\label{app:scalar_field}

Finally, we can simplify the field equations~(\ref{eq:low_field2}) with a weak field expansion for the scalar field around $\phi_\infty$ by expanding the conformal factor 
\begin{equation}\label{eq:exp_conf}
\frac{A_l(\varphi)}{A_{l,\infty}}\approx 1 + \alpha_{l,\infty}(\varphi-\varphi_\infty),
\end{equation}
where $A_{l,\infty}=A_l(\varphi_\infty)$. This expansion is valid if $ \alpha_{l,\infty}(\varphi-\varphi_\infty) \ll 1$ or using the scalar field solution~(\ref{eq:scal_sol}) if $\alpha_{l,\infty}^2$ and $\alpha_{l,\infty}\alpha_{d,\infty}$ are not too large. A more accurate estimation with the parameters used for the galaxies considered in this study shows that:
\begin{subequations}\label{eq:low_scal}
	\begin{eqnarray}
		\alpha_{l,\infty}^2 &< &10^3 \, ,\\
		\alpha_{l,\infty}\alpha_{d,\infty} &<& 10^3 \, .
	\end{eqnarray}
\end{subequations}
This assumption can be justified by a naturalness argument or can be justified a posteriori by our results (see the discussion in Sec.~\ref{sec:model}). We need to introduce the expression of the matter density used. The density profiles we consider are discussed in Sec.~\ref{sec:model} and have the form (\ref{eq:DensTot}) and (\ref{eq:DensLum})
\begin{align}
\rho_{t}(r)&= \rho_{l}(r) + \rho_{d}(r) =  r^{ - \mathbf{\gamma}} {\rho^{(\gamma)}_{0}} \, ,\qquad \mbox{and} \qquad  \rho_{l}(r)= r^{ - \mathbf{\delta}} {\rho^{(\mathbf{\delta)}}_{0}}\, ,
\end{align}
where $\gamma$ and $\delta$ are two dimensionless constants and $\rho^{(\gamma)}_{0}$ and $\rho^{(\delta)}_{0}$ are constants with the appropriate dimension.

Using these expressions for the density in Eq.~(\ref{eq:scalar_low_field}) leads to
\begin{align}
\ddot{\varphi} + \frac{2}{r} \dot{\varphi} &=- \frac{4 \pi G}{c^2} \left( \alpha_{l, \infty} \rho_{l} +  \alpha_{d, \infty}\rho_{d} \right)= - \frac{k_\gamma}{r^\gamma} -  \frac{k_\delta}{r^\delta} 
\end{align}
where we introduce the constants
\begin{subequations}\label{eq:ks}
	\begin{eqnarray}
		k_{\delta} &=&  \frac{4 \pi G}{c^2} \left( \alpha_{l, \infty} - \alpha_{d, \infty} \right) \rho^{(\delta)}_0		\, ,\\
		k_{\gamma} &=& \frac{4 \pi G}{c^2}\alpha_{d, \infty}\rho^{(\gamma)}_0 .
	\end{eqnarray}
\end{subequations}
The scalar field equation has a general solution of the form $\varphi(r) =\varphi_\infty+ \varphi_\gamma(r) + \varphi_\delta(r) $, where
\begin{align} \label{eq:scal_sol}
\varphi_{j}(r) &=
		\frac{k_j}{\left(j - 3\right) \left(j - 2\right)} r^{2- j}\, , \qquad \textrm{for} \qquad j=\gamma \textrm{ or } \delta\, .
\end{align}

\section{Gravitational lensing}\label{app:lensing}
\subsection{Deflection angle} \label{def_App}
The deflection angle is denoted by $\widehat{\vec{\alpha}}$. The quantity $\widehat{\vec{\alpha}}$, and others with arrows are two dimensional vectors. In principle, an observer will observe a two dimensional image, and quantify the bending along two axes. Under the additional assumption that the deflection angle is small, $\widehat{\vec \alpha}$ can be computed as an integral of the gradient of the sum of the two gravitational potentials over the unperturbed line of sight in comoving coordinates. This is known as the Born approximation, and is valid for galaxies and clusters \cite{primod}. It is common to introduce the rescaled deflection angle $\alpha$ defined by:
\begin{align}
\vec{\alpha} &= \frac{D_{ls}}{D_s} \widehat{\vec{\alpha}} \, ,
\end{align}
where $D_{ls}$ is the angular distance between the lens and source and $D_s$ is the angular distance to the source. This angle corresponds to the observed gravitational deflection angle so that the lens equation simply takes the form:
\begin{align} \label{lens_eq}
\vec{\theta} &= \vec{\theta}_S + \vec{\alpha } \, ,
\end{align}
where $\vec{\theta}$ is the observed angular position of the image and $\vec{\theta}_S$ is the angular position of the source in the absence of the lens.

Under the assumption that the deflection angle is small \cite{lens_schn}, the angle $\vec \alpha$ can be computed as
\begin{align}
 \vec{\alpha}&= \frac{D_{ls}}{D_s c^2} \int \nabla_{\perp}  \left( {\Phi}\left(r\right) + {\Psi}\left(r \right) \right)\  dD ,
\end{align}
where $\nabla_\perp$ is the two dimensional gradient perpendicular to the line of sight. The previous integral is performed along a line parallel to the line of sight. The variable $D$ denotes the projection of the coordinates along the line of sight and $\xi$ is the radius of the projection on a plane orthogonal to the line of sight (so that $r^2=D^2+\xi^2$). The angle $\theta$ is defined as the angular coordinate on the plane and is related to the $\xi$ coordinate by $\xi=\theta D_l$. 

Taking the 2-D angular gradient of the previous quantity leads to
\begin{align}
\nabla_\theta \cdot  \vec{\alpha} &= \frac{D_{ls}}{D_s D_lc^2} \int  \Delta \left( {\Phi} + {\Psi} \right)\  dD.
\end{align} 
As a result of Eq.~(\ref{laplacian_pot}), we can relate the deflection to the matter density of the lens
\begin{align}\label{eq:lens_surfDens}
\nabla_\theta \cdot  \vec{\alpha} &= \frac{D_{ls}}{D_s D_lc^2} \int  8\pi G\frac{A^4_l(\varphi)}{A^4_{l,\infty}}\rho_t(r) \  dD =\frac{2 \Sigma}{ \Sigma_c} \, ,
\end{align}
where we introduce the surface mass density
\begin{align} \label{eq:surf_dens}
\Sigma \left(\xi \right) &=  \int_{-\infty}^\infty \frac{A^4_l(\varphi)}{A^4_{l,\infty}}\rho_t(r) \  dD,
\end{align}
and the critical surface mass density
\begin{align}
\Sigma_c &= \frac{c^2}{4 \pi G} \frac{D_s}{D_{ls} D_{l}}\, .
\end{align}

A lens is fully characterized by its surface density, and the degree of image distortion depends on $\frac{\Sigma}{\Sigma_{c}}$ \cite{1996astro.ph..6001N}. Under the assumptions~(\ref{eq:low_scal}), we can expand the conformal factor using~(\ref{eq:exp_conf}) in the surface density Eq.~(\ref{eq:surf_dens_tot}) which leads to
\begin{equation}\label{eq:sigma_dec}
	\Sigma \left( \xi\right) =\Sigma_\textrm{GR}\left( \xi\right) +\Sigma_\varphi\left( \xi\right)=\int_{-\infty}^\infty \rho_t\left(r\right) \  dD + 4 \alpha_{l,\infty}\int_{-\infty}^\infty (\varphi\left(r\right)-\varphi_\infty)\rho_t\left(r\right) \  dD \, ,
\end{equation}
where $r=\sqrt{\xi^2+D^2}$ and where we introduce $\Sigma_\textrm{GR}$ the expression of the surface density in GR and $\Sigma_\varphi$ the correction due to the modification of the theory of gravity considered. Making use of the scalar field solution obtained in Eq.~(\ref{eq:scal_sol}), we can compute the GR and scalar field contributions to the surface density in our model as
\begin{subequations} \label{eq:surf_dens_tot}
\begin{align}
\Sigma_\textrm{GR}  \left(\xi\right)&= \sqrt{\pi } \xi ^{1-\gamma } \lambda(\gamma) \rho^{(\gamma)}_0 \, ,\\
\Sigma_{\varphi} \left( \xi \right) &=  4 \alpha_{l, \infty} \rho_0^{(\gamma)} \sqrt{\pi}  \left( k_\gamma\xi ^{3-2 \gamma }  \frac{ \lambda (2 \gamma -2)}{\left( \gamma - 3 \right) \left( \gamma - 2\right)} + k_\delta\xi ^{3-\gamma -\delta } \frac{ \lambda (\gamma +\delta -2) }{\left(\delta - 3 \right) \left( \delta - 2 \right)}  \right) \, ,
\end{align}
\end{subequations}
where  
\begin{equation}
\lambda(\gamma) = \frac{\Gamma\left(\frac{\gamma - 1}{2} \right)}{\Gamma\left(\frac{\gamma}{2} \right)}\, .
\end{equation}
The total projected mass within a 2D disk of radius $\theta$ in the sky will be denoted by $\bar M(\theta)$ whose expression is given by
\begin{align}
\bar M\left( \theta \right) &=\int_0^{\xi=\theta D_l} 2 \pi x \Sigma \left( x \right) \ d x= D_l^2\int_0^\theta 2 \pi \theta' \Sigma \left( \theta' D_l \right) \ d \theta'\nonumber \\
&=\bar M_\textrm{GR}\left( \theta \right) +\bar  M_\varphi\left( \theta \right)  =  D_l^2\int_0^\theta 2 \pi \theta' \Sigma_\textrm{GR} \left( \theta' D_l \right) \ d \theta' +D_l^2 \int_0^\theta 2 \pi \theta' \Sigma_\varphi \left( \theta' D_l \right) \ d \theta.' \label{eq:barm}
\end{align}
The GR and scalar field contributions are then given by the expressions:
\begin{subequations} \label{Mass_eq}
\begin{align}
\bar M_\textrm{GR} \left( \theta \right) &= \frac{2 \pi^{\frac{3}{2}} (\theta D_l)^{3-\gamma} \rho_0^{(\gamma)} \lambda (\gamma )  }{(3 - \gamma)}  \, ,\\
 \bar M_\varphi \left( \theta \right) &= 4\alpha_{l,\infty}\frac{3-\gamma}{\lambda(\gamma)}\bar M_\textrm{GR}(\theta)\left( k_\gamma\frac{\left(D_l \theta \right)^{2- \gamma} \lambda\left(2\gamma -2 \right) }{ \left( \gamma -3 \right) \left( \gamma -2 \right) \left( 5 - 2\gamma \right)} + k_\delta \frac{\left(D_l \theta \right)^{2- \delta} \lambda\left(\gamma + \delta -2 \right) }{ \left( \delta -3 \right) \left( \delta -2 \right) \left( 5 - \gamma - \delta \right)}\right)\, . 
\end{align}
\end{subequations}
 In cylindrical coordinates, and for an axially symmetric lens, Eq.~(\ref{eq:lens_surfDens}) yields
\begin{align}\label{eq:alpha_th}
 \alpha\left( \theta \right) &=  \frac{2}{\Sigma_c \theta} \int_0^{\theta} \theta' \Sigma \left( \theta' D_l \right) \ d \theta' =\frac{\bar M(\theta)}{\Sigma_c \theta \pi D_l^2} = \frac{D_{ls}}{D_l D_s}\frac{4G\bar M(\theta)}{\theta}\, .
\end{align}

Using the expression for the surface density given by Eq.~(\ref{eq:surf_dens_tot}) and introducing the expansion of the gravitational deflection angle
\begin{equation}\label{eq:alpha_exp}
	\alpha \left(\theta \right)=\alpha_\textrm{GR}\left(\theta \right) + \alpha_{\varphi}\left(\theta \right)\, ,
\end{equation}
where $\alpha_\textrm{GR}\left(\theta \right)$ is the standard GR expression and $\alpha_\varphi\left(\theta \right)$ is the correction due to the modification of the theory of gravity. Using Eqs.~(\ref{eq:alpha_th}) and (\ref{Mass_eq}), one gets
\begin{subequations}  \label{defl}
\begin{align}
 \alpha_\textrm{GR} \left(\theta \right)&= \frac{D_{ls}}{D_l D_s}\frac{4  G \bar M_\textrm{GR}\left( \theta\right)}{\theta }=\frac{\bar M_\textrm{GR}(\theta)}{\theta D_l^2 \pi \Sigma_c} \, ,\\
\alpha_{\varphi} \left(\theta \right)&= \frac{D_{ls}}{D_l D_s}\frac{4  G \bar M_{\varphi}\left( \theta\right) }{\theta}=\frac{\bar M_\varphi(\theta)}{\theta D_l^2 \pi \Sigma_c} \, .
\end{align}
\end{subequations}

\subsection{Einstein radius} \label{EinR_App}
An Einstein ring is observed when the lens is axially symmetric, and the source, lens and observer are aligned. In this case, $\theta_S = 0$ in Eq.~(\ref{lens_eq}). The angular radius of the Einstein ring is called the Einstein radius. It is the angle $\theta_E$ such that
\begin{align}\label{eq:einstein_def}
\alpha \left( \theta_E \right) &= \theta_E.
\end{align}
We consider the scalar field contribution to the Einstein radius as a first order perturbative correction to the GR value
\begin{equation}
	\theta_E=\tegr+\theta_{E,\varphi}\, .
\end{equation}
Introducing this expansion and using the expansion (\ref{eq:alpha_exp}) in the Eq.~(\ref{eq:einstein_def}) leads to
\begin{equation}
	\tegr+\theta_{E,\varphi}=\alpha_\textrm{GR}(\tegr)+\alpha_\varphi(\tegr)+\left.\frac{\partial \alpha_\textrm{GR}}{\partial \theta}\right|_{\theta=\tegr}\theta_{E,\varphi}\, .
\end{equation}
Solving the previous equation order by order and using the expression for $\alpha$ given by Eq.~(\ref{defl}) leads to 
\begin{subequations}
	\begin{align}
		\tegr&=\alpha_\textrm{GR}(\tegr)\, ,\\
	\theta_{E,\varphi} &= \frac{\alpha_\varphi(\tegr)}{1-\left.\frac{\partial \alpha_\textrm{GR}}{\partial \theta}\right|_{\theta=\tegr}}= \frac{\alpha_{\varphi}\left(\tegr\right)}{\gamma - 1} \, .
	\end{align}	
\end{subequations}
Introducing the expression of $\alpha$ given by the relations (\ref{defl}) in the previous expressions finally leads to:
\begin{subequations} \label{eq:ein_rad}
\begin{align}
\tegr &= \left(\frac{D_{ls}}{D_s D_l} 4 G M_\textrm{GR}(\tegr) \right)^{\frac{1}{2}}=\left(2\frac{\sqrt{\pi}\rho_0^{(\gamma)}\lambda(\gamma)}{\Sigma_c (3-\gamma)}\right)^{\frac{1}{\gamma-1}} \frac{1}{D_l} \, ,\\
\theta_{E,\varphi} &= \frac{\tegr \bar M_\varphi(\tegr)}{(\gamma-1)\bar M_\textrm{GR}(\tegr)}\\
&=4\alpha_{l,\infty}\frac{3-\gamma}{\gamma-1}\frac{\tegr}{\lambda(\gamma)}\left( k_\gamma\frac{\left(D_l \tegr \right)^{2- \gamma} \lambda\left(2\gamma -2 \right) }{ \left( \gamma -3 \right) \left( \gamma -2 \right) \left( 5 - 2\gamma \right)} + k_\delta \frac{\left(D_l \tegr \right)^{2- \delta} \lambda\left(\gamma + \delta -2 \right) }{ \left( \delta -3 \right) \left( \delta -2 \right) \left( 5 - \gamma - \delta \right)}\right)\, .
\end{align}
\end{subequations}

\section{Velocity dispersion} \label{app:vel_disp}
The line-of-sight stellar velocity dispersion, $\sigma_r^2$, is dependent on the total density (through the gravitational potential), and the luminosity density \cite{2006ApJ...649..599K}. The spherical Jeans equations allow us to solve for the line-of-sight stellar velocity dispersion as a function of the radius \cite{2008gady.book.....B}:
\begin{align}
\frac{\partial ( \rho_{l} \sigma_r^2 )}{\partial r}+ \frac{2 \beta}{r} \rho_{l} \sigma_r^2 &= - \rho_{l} \frac{\partial \Psi}{\partial r}\, , \label{eq:eq_mot_vel_disp}
\end{align}
where $\Psi$ is the standard gravitational potential which enters the time component of the Jordan frame metric tensor (\ref{eq:metric_WJF}), $\sigma_r$ is the radial component of the velocity dispersion and  $\beta = 1 - \frac{\sigma^2_t}{\sigma_r^2}$ is the velocity anisotropy parameter with $\sigma_t$ the tangential component of the velocity dispersion. All the previous quantities depend on the radial coordinate. Assuming that $\beta$ is constant (\citep{2001AJ....121.1936G, 2006ApJ...649..599K}), the solution to the above equation is
\begin{align}
\sigma_r^2 \left( r \right) &=\frac{1}{\rho_{l}(r) r^{2 \beta}} \int^\infty_r  \rho_{l}(x) x^{2 \beta} \dot \Psi(x) dx \, ,
\end{align}
where the expression of $\Psi$ is given by Eq.~(\ref{Psi_eq}). As usual, we decompose $\Psi$ into a GR part $\Psi_\textrm{GR}$ and a correction $\Psi_\varphi$. Integrating the Eq.~(\ref{eq:nu_low_field}) gives the GR value
\begin{subequations}
\begin{align}
\dot{\Psi}_\textrm{GR} &= \frac{G M_\textrm{GR}(r)}{r^2}=\frac{ 4 \pi G\int_0^r x^2 \rho_t(x) \ dx}{r^2}\,.
\end{align}
Note that $M(r)$, the total mass included in a volume of radius $r$, should no be confused with the mass within a 2D disk of radius $\theta$ on the sky $\bar M(\theta)$ defined by Eq.~(\ref{eq:barm}).

Moreover, the correction $\Psi_\varphi$ is given by the second term in Eq.~(\ref{Psi_eq}), so that
\begin{align}
	\dot{\Psi}_{\varphi} &=c^2 \alpha_{l, \infty} \dot \varphi\, .
\end{align}
\end{subequations}
Using the previous expressions leads to 
\begin{subequations}
\begin{align}
	\sigma_r^2 \left( r \right) &=\sigma_{r,GR}^2 \left( r \right)+\sigma_{r,\varphi}^2 \left( r \right)\label{eq:sigma_sum}\\
	&=\frac{G}{\rho_{l}(r) r^{2 \beta}} \int^\infty_r \rho_{l} (x) x^{2 \beta-2} M(x) dx +\frac{c^2 \alpha_{l,\infty}}{\rho_{l} (r) r^{2 \beta}} \int^\infty_r \rho_{l} (x) x^{2 \beta} \dot \varphi(x) dx \, ,
\end{align}
\end{subequations}
with the expression of $\varphi$ given by Eq.~(\ref{eq:scal_sol}).

The actual observable from velocity dispersion data is the luminosity weighted over the line of sight and over the effective spectrometer aperture, which is described by the aperture weighting function
\begin{align}
w\left( R \right) = e^{\frac{-R^2}{2 \bar{\sigma}_{atm}^2}},
\end{align}
where the standard atmospheric seeing $\bar{\sigma}_{atm}$ is computed in \cite{2006ApJ...638..703B,2008ApJ...682..964B} for the dataset considered in this paper.
The observable is given by the expression \cite{bolton:2006uq,schwab:2010fk}:
\begin{align}
\sigma^2_{\star} &= \frac{\int_0^\infty dR \, R \, w(R) \int_{-\infty}^\infty dz \rho_{l}(r) \left( 1 - \beta  \frac{R^2}{r^2 }\right) \sigma_r^2(r)}{\int_0^\infty dR \, R\, w(R)\int_{-\infty}^\infty dz \rho_{l} (r)} = \sigma^2_{\star,GR } + \sigma_{\star,\varphi }^2 \, ,
\end{align}
where $r^2 = R^2 + z^2$ and where we introduce the decomposition (\ref{eq:sigma_sum}).

We write down the two contributions to the velocity dispersion below. The GR contribution is:
\begin{subequations}\label{eq:vel_disp_gr}
\begin{align} 
\sigma^2_{\star,GR } &= \rho_0^{(\gamma)} j,
\end{align}
where the coefficients $j$ reads:
\begin{align}
j = \frac{ 4 {\pi} G (2 \bar{\sigma}_{atm}^2) ^{1-\gamma/2 }}{(3-\gamma)(\gamma+\delta-2\beta-2)}\,  \frac{ \Gamma \left(\frac{5-\gamma-\delta}{2} \right)}{\Gamma \left(\frac{ 3 - \delta}{2}\right)} \, \frac{  \lambda\left(  \gamma +\delta -2  \right)  -  \beta \lambda\left( \gamma +\delta   \right)    }{ \lambda \left( \delta \right) }.
\end{align}
\end{subequations}
The scalar field contribution is:
\begin{subequations}\label{eq:vel_disp_phi}
\begin{align} 
\sigma_{\star,\varphi }^2 &=  \rho_0^{(\gamma)} m_1 + \rho_0^{(\delta)} m_2,
\end{align}
where the expressions of the two constants are given by
\begin{align}
m_1 =   \frac{4 \pi G \alpha^2_{l, \infty}\left( 2\bar{\sigma}^2_{atm}\right) ^{1 -\gamma/2}  }{\left( 3-\gamma  \right) \left( \gamma +\delta - 2\beta -2 \right)} \, \frac{\Gamma \left(\frac{5-\gamma -\delta}{2}\right)}{\Gamma \left(\frac{3 - \delta}{2}\right)} \,  \frac{\lambda(\gamma + \delta -2) -\beta\lambda \left(\gamma +\delta\right)}{\lambda \left( \delta \right)} =j\alpha_{l,\infty}^2\, ,
\end{align}
and
\begin{align}
m_2 =    \frac{4 \pi G \alpha_{l, \infty} \left( \alpha_{l, \infty} - \alpha_{d, \infty} \right) \left( 2\bar{\sigma}^2_{atm} \right) ^{1- \delta/2 } }{\left(3-\delta  \right)\left( 2\delta - 2\beta-2 \right)} \,  \frac{\Gamma \left(\frac{5}{2}-\delta \right)}{\Gamma \left(\frac{3 - \delta}{2} \right)}\,  \frac{\lambda (2 \delta - 2) - \beta \lambda \left( 2\delta \right) }{\lambda \left( \delta \right) }  \, .
\end{align}
\end{subequations}

\section{Density parameters} \label{DensPar_App}
In our model the variable of statistical interest is the total velocity dispersion $\sigma_{\star}$ (see Sec.~\ref{sec:data}). It is therefore necessary to express it as a function of the variables that will be used in the Monte Carlo (i.e. $\alpha_{l,\infty}$, $\alpha_{d,\infty}$, $\gamma$ and $\beta$). The GR and scalar field contributions to the Einstein radius and velocity dispersion computed in sections \ref{EinR_App} and \ref{app:vel_disp} have a phenomenological dependence on the density parameters $\rho^{(\delta)}_{0}$ and $\rho^{(\gamma)}_{0}$ that need to be removed. As usual these parameters can be decomposed into a GR contribution and a scalar field correction
\begin{subequations} \label{rho_pert}
\begin{align}
\rho^{(\delta)}_{0} &= \rho^{(\delta)}_{0, GR} + \rho^{(\delta)}_{0, \varphi}  \, ,\\
\rho^{(\gamma)}_{0} &= \rho^{(\gamma)}_{0, GR} + \rho^{(\gamma)}_{0, \varphi} \, .
\end{align}
\end{subequations}

In Sec.~\ref{lum_dens_app}, we show how the knowledge of the SM (or stellar) mass $M_*$ (see Sec.~\ref{sec:velocity_disp} to see how this variable can be obtained) can be used to determine $\rho^{(\delta)}_{0, \textrm{GR}}$ and $\rho_{0,\varphi}^{(\delta)}$. Then, the observation of the Einstein radius allows us to determine $\rho^{(\gamma)}_{0,\textrm{GR}}$ and $\rho^{(\gamma)}_{0, \varphi}$ as we will show in Sec.~\ref{tot_dens_app}. As a result we are able to predict the total velocity dispersion depending on the free parameters $\gamma$, $\beta$, $\alpha_{l,\infty}$, $\alpha_{d,\infty}$ together with observations of the Einstein radius $\theta_*$ and the value of the SM mass $M_*$.

\subsection{Luminous matter density parameter} \label{lum_dens_app}
The density parameters $\rho^{(\delta)}_{0, \textrm{GR}}$ and $\rho_{0,\varphi}^{(\delta)}$ can be determined through the perturbation of the SM mass. Given the SM surface density of a galaxy, $\Sigma_\textrm{lum}$, the SM mass contained within some de-projected radius $\xi$ is given by
\begin{align}
\bar M_\textrm{lum} \left( \xi \right) &= 2 \pi \int_0^{\xi} \Sigma_\textrm{lum} \left( \xi' \right) \xi' \ d\xi'.
\end{align}
The SM surface density of a galaxy is defined similarly to the total surface mass density defined by Eq.~(\ref{eq:surf_dens}):
\begin{equation}
 \Sigma_\textrm{lum}(\xi)=\int_{-\infty}^\infty \frac{A^4_l(\varphi)}{A^4_{l,\infty}} \rho_l(r)dD =\int_{-\infty}^\infty \rho_l\left(r\right) \  dD + 4 \alpha_{l,\infty}\int_{-\infty}^\infty (\varphi\left(r\right)-\varphi_\infty)\rho_l\left(r\right) \  dD\, ,
\end{equation}
where the radial coordinate is given by $r=\sqrt{\xi^2+D^2}$. 

The calculation of the SM mass is very similar to the calculation of the projected total mass that leads to Eqs.~(\ref{Mass_eq}). In the SM case, the equations become:
\begin{subequations} \label{lum_mass}
\begin{align}
\bar M_\textrm{lum,GR}  &= \frac{2 \pi^{\frac{3}{2}} R_g^{3-\delta} \rho_0^{(\delta)} \lambda (\delta)}{3 - \delta}  \, ,\\
 \bar M_{\textrm{lum},\varphi} \left( \theta \right) &= 4\alpha_{l,\infty}\frac{3-\delta}{\lambda(\delta)}\bar M_\textrm{lum,GR}\left( k_\gamma\frac{R_g^{2- \gamma} \lambda\left(\gamma+\delta -2 \right) }{ \left( \gamma -3 \right) \left( \gamma -2 \right) \left( 5 - \gamma-\delta \right)} + k_\delta \frac{R_g^{2- \delta} \lambda\left(2 \delta -2 \right) }{ \left( \delta -3 \right) \left( \delta -2 \right) \left( 5 -2 \delta \right)}\right)\, , \label{eq:mlumphi}
\end{align}
\end{subequations}
where $R_g$ is the observed radius of the galaxy.

By equating the expression of the SM mass with its observed value $M_*$ leads to
\begin{equation}
M_{*} =  \bar M_\textrm{lum,GR} \left( \rho^{(\delta)}_{0, GR} \right) + \left. \frac{d \bar M_\textrm{lum,GR}}{d \rho^{(\delta)}_{0} }\right|_{ \rho^{(\delta)}_{0, \textrm{GR}}}   \rho^{(\delta)}_{0,\varphi} + \bar M_{\textrm{lum},\varphi} \left( \rho^{(\delta)}_{0, GR} \right) \, .
\end{equation}
The last expression can be inverted order by order by using Eqs.~(\ref{lum_mass}) to find the density parameter $\rho_0^{(\delta)}$:
\begin{subequations} \label{rho_lum_exact}
\begin{align}
\rho^{(\delta)}_{0,\textrm{GR}}  &= \frac{M_{*} (3 - \delta)}{2 \pi^{3/2}\lambda(\delta) R_g^{(3 - \delta)}}\, , \label{eq:rho_d_GR}\\
\rho^{(\delta)}_{0, \varphi} &= -\frac{\bar M_{\textrm{lum},\varphi} \left( \rho^{(\delta)}_{0, GR} \right)}{\left. \frac{d \bar M_\textrm{lum,GR}}{d \rho^{(\delta)}_{0} }\right|_{ \rho^{(\delta)}_{0, \textrm{GR}}}}=-\frac{ \rho^{(\delta)}_{0, GR}}{M_{*}} \bar M_{\textrm{lum},\varphi} \left( \rho^{(\delta)}_{0, GR} \right)\, . \label{eq:rho_d_phi}
\end{align}
\end{subequations}

\subsection{Total matter density parameter}\label{tot_dens_app}
Here, we will show how the observed Einstein radius $\theta_*$ can be used to infer the value of $\rho_0^{(\gamma)}$. In Sec.~\ref{EinR_App}, we derived the GR and scalar field contributions to the Einstein radius $\theta_E$. Both contributions are dependent on the density parameters $\rho^{(\gamma)}_{0}$ and $\rho^{(\delta)}_{0}$. The value of $\rho^{(\delta)}_{0}$ is supposed to be known from the luminosity mass (see the previous section). We use the usual decomposition~(\ref{rho_pert}) and expand to first order to obtain:
\begin{equation}
 \theta_*= \theta_E\left(\rho_0^{(\gamma)}\right)=\tegr\left(\rho_{0,\textrm{GR}}^{(\gamma)}\right) + \left.\frac{d \tegr}{d\rho^{(\gamma)}_0}\right|_{\rho_{0,\textrm{GR}}^{(\gamma)}}\rho_{0,\varphi}^{(\gamma)} + \theta_{E,\varphi}\left(\rho_{0,\textrm{GR}}^{(\gamma)}\right)\, .
\end{equation}
Using Eqs.~(\ref{eq:ein_rad}) to determine $\rho^{(\gamma)}_{0, GR}$ and $\rho^{(\gamma)}_{0, \varphi}$, one gets:
\begin{subequations}\label{eq:rhogamma}
\begin{align}
\rho^{(\gamma)}_{0,\textrm{GR}} &= \frac{\left( D_l \theta_*  \right)^{\gamma - 1 }}{2 \sqrt{\pi} \lambda(\gamma)} \Sigma_c \left( 3 - \gamma \right)\, , \label{eq:rho_g_GR}\\
\rho^{(\gamma)}_{0, \varphi} &=-\frac{\theta_{E,\varphi}\left(\rho_{0,\textrm{GR}}^{(\gamma)}\right)}{\left.\frac{d \tegr}{d\rho^{(\gamma)}_0}\right|_{\rho_{0,\textrm{GR}}^{(\gamma)}}} =- \frac{\rho^{(\gamma)}_{0, GR}}{\theta_*} \left( \gamma - 1 \right)  \theta_{E, \varphi}\left( \rho^{(\gamma)}_{0, GR} \right) \\
&=- 4 \alpha_{l, \infty}\rho^{(\gamma)}_{0,\textrm{GR}} \frac{(3 - \gamma)}{\lambda(\gamma)}\left[k_{\gamma} \frac{ ( D_l\theta_*)^{2 - \gamma} \lambda\left(2 \gamma - 2 \right) }{(5 - 2 \gamma)(\gamma - 3)(\gamma -2) } + k_\delta \frac{(\theta_*D_l)^{2 - \delta}\lambda (\gamma +\delta -2)}{5 - \gamma - \delta)(\delta - 3 )(\delta -2 )} \right] \, , \label{eq:rho_g_phi}
\end{align}
\end{subequations}
where in the last expression, one uses the GR values for the density parameter in the coefficients $k_\gamma$ and $k_\delta$ given by Eqs.~(\ref{eq:ks}). Therefore, it is clear that the scalar field contribution to the density parameter $\rho^{(\gamma)}_{0, \varphi}$ explicitly depends on the SM mass density parameter $ \rho^{(\delta)}_{0}$ through the scalar field solution. This dependence presents no problems as Eqs.~(\ref{rho_lum_exact}) allow us to specify this contribution to $\rho^{(\gamma)}_{0}$.

\bibliographystyle{JHEP}
\bibliography{mybib}

\end{document}